\documentclass[pre,twocolumn,amsmath,amssymb,showpacs]{revtex4}
\usepackage{amsmath,amssymb,latexsym}
\usepackage{graphicx,epstopdf}
\allowdisplaybreaks[1]
\newcommand{\bs}[1]{\boldsymbol{#1}}
\newcommand{\ca}[1]{\mathcal{#1}}
\newcommand{\bfq}{\mbox{\boldmath $\it q$}}
\newcommand{\bfp}{\mbox{\boldmath $\it p$}}
\newcommand{\bfGamma}{\mbox{\boldmath $\it\Gamma$}}
\newcommand{\bfsGamma}{\mbox{\boldmath $\scriptstyle\it\Gamma$}}

\newcommand{\Cd}{C}

\newcommand{\spaEq}{\hspace{2ex}}
\newcommand{\widthfig}{0.45\textwidth}
\newcommand{\widthfigB}{0.40\textwidth}
\newcommand{\vspfigA}{\vspace{3ex}}  
\newcommand{\vspfigB}{\vspace{0ex}} 
\newcommand{\vspfigC}{\vspace{3ex}}
\newcommand{\sumtwo}[2] 
   { \underset{(j>k)}{\sum^{\ca{N}}_{#1}\sum^{\ca{N}}_{#2}}\;}  
\newcommand{\prodtwo}[2] 
   { \underset{(j>k)}{\prod^{\ca{N}}_{#1}\prod^{\ca{N}}_{#2}}\;}  

\begin{document}
%
\title{Stochastic boundary approaches to many-particle systems coupled to a particle reservoir}
\author{Tooru Taniguchi and  Shin-ichi Sawada}
\affiliation{School of Science and Technology, Kwansei Gakuin University, 2-1 Gakuen, Sanda, Hyogo, Japan} 
\date{\today}
\begin{abstract} 
Stochastic boundary conditions for interactions with a particle reservoir are discussed in many-particle systems. 
We introduce the boundary conditions with the injection rate and the momentum distribution of particles coming from a particle reservoir in terms of the pressure and the temperature of the reservoir. 
It is shown that equilibrium ideal gases and hard-disk systems with these boundary conditions reproduce statistical-mechanical properties based on the corresponding grand canonical distributions.
We also apply the stochastic boundary conditions to a hard-disk model with a steady particle current escaping from a particle reservoir in an open tube, and discuss its nonequilibrium properties such as a chemical potential dependence of the current and deviations from the local equilibrium hypothesis.    

\end{abstract}
   
\pacs{
05.20.-y, 
02.50.Ey, 
51.10.+y 
}

\maketitle  
\section{Introduction} 
\label{Introduction} 

Systems interacting with a thermal reservoir are frequently used in descriptions of a variety of physical phenomena. 
For such systems, the reservoir plays roles not only to keep systems in an equilibrium state, but also to make systems relax to an equilibrium state, and even to cause nonequilibrium steady states by interactions with different reservoirs. 
Depending on the type of reservoirs, various types of thermodynamic quantities are kept to be fixed in the systems, for example, a temperature (by the heat reservoir), a chemical potential (by the particle reservoir) and a pressure.   
Because of the importance of interactions with a thermal reservoir in the foundation and applications of thermodynamics and statistical mechanics, many effective dynamical approaches to interactions with a thermal reservoir have been proposed, such as the Langevin or the Fokker-Planck equations \cite{R89,K92,CK12}, Hamiltonian dynamics with linear couplings with infinite number of harmonic oscillators (e.g. the Caldeira-Leggett model) \cite{P10,W08}, the deterministic thermostat dynamics (e.g. the Nose-Hoover thermostat and the isokinetic thermostat) \cite{CP91,T99,FS02,EM08,T20,LM15}, the Anderson thermostat \cite{T99,FS02,A80} and the local Monte Carlo method \cite{PB93,HS94}. 
Many efforts have been done to check justifications of these approaches as thermostats, via calculations of various statistical-mechanical properties such as the ergodicity and momentum distributions, leading to their improvements \cite{FS02,T20,LM15}. 

The stochastic boundary approach is a dynamical method to take into account of effects of interactions with a thermal reservoir as  boundary conditions involving stochastic processes.  
In this approach, for example, interactions of a system with a thermal reservoir on a boundary $\ca{S}$ are described by the boundary condition on $\ca{S}$ in which the momentum distribution of particles coming from the boundary $\ca{S}$ is given by the equilibrium distribution corresponding to the temperature of the heat reservoir \cite{CT80,TC82,MK84,MK87,BL91,TT98}. 
Systems with such stochastic boundary conditions for interactions with a single heat reservoir relax to an equilibrium state with the temperature of the reservoir, and dissipations in the systems occur on the boundary $\ca{S}$. 
In systems with the stochastic boundary conditions, some statistical-mechanical features, such as consistencies with the kinetic theory \cite{MK87,LS78}, the local equilibrium hypothesis \cite{TC82,MK87,BL91}, the specific heat \cite{CT80} and spatial energy distributions \cite{TT98}, have been investigated.  
In the sense of using boundary conditions to describe interactions with a thermal reservoir, the methods using the master equation \cite{BL55,LF57,LS78}, numerical evaluations of boundary conditions \cite{PL12} or quantum scattering techniques \cite{D95,I97,LJ13}, etc., are also regarded as stochastic boundary approaches.    
An advantage of these boundary methods is that in spite of describing interactions with a thermal reservoir, they do not need to introduce explicitly degrees of freedom of particles inside the reservoir. 
Moreover, these methods can be used to discuss nonequilibrium properties and phenomena, such as Onsager's reciprocal relations \cite{BL55}, heat currents \cite{CT80,TC82,MK84,MK87,BL91,BL55,LF57}, electric currents \cite{D95,I97,LJ13} and thermoelectric effects \cite{LJ13,CM08,HP09}.

The principal aim of this paper is to discuss the stochastic boundary conditions for many-particle systems coupled to a particle reservoir with particle-particle interactions. 
In order to describe effects of interactions with a particle reservoir, we introduce the particle injection rate to the systems from the reservoir via the boundary $\ca{S}$ with the pressure of the reservoir which specifies a momentum transport via $\ca{S}$, while the momentum distribution of injected particles on $\ca{S}$ is given by the equilibrium distribution corresponding to the temperature of the reservoir. 
Since the pressure and the momentum distribution of the injected particles on $\ca{S}$ are given by the temperature $T$ and the chemical potential $\mu$ of the particle reservoir, our particle-injecting stochastic boundary (PISB) conditions are specified by these thermodynamic quantities $T$ and $\mu$ as external parameters. 
This method gives a simple and effective way of describing many-particle systems coupled to a particle reservoir by boundary conditions, differently from other reservoir-coupling approaches \cite{PB93,HS94,WS12} including the dynamics of particles inside a particle reservoir.

As applications of this PISB formulation to equilibrium many-particle systems, we consider ideal gases and many-hard-disk systems. 
Because of a simplicity of hard-core interactions, hard-disk (or hard-sphere) systems, as typical examples of many-particle systems with particle-particle interactions, have been widely used to investigate statistical-mechanical properties by numerical methods \cite{AT87,H92} and analytical methods \cite{GG84,M08,HM13,S13}.  
Since equilibrium statistical-mechanical quantities not only of ideal gases but also of many-hard-disk systems in low-density cases can be calculated analytically based on the grand canonical distribution, we use these systems to show that the systems with the PISB conditions can reproduce correctly equilibrium properties of the systems coupled to a particle reservoir.  
Furthermore, we apply the PISB method to particle-escape phenomena \cite{BB90,TS11a,AP13,TS14}, as a typical example of nonequilibrium phenomena. 
We consider the systems in which hard disks escaping from a particle reservoir into one end of a tube leave from another end of the tube, and produce a steady particle current inside the tube after a long time. 
By using the PISB approach, we analyze statistical-mechanical properties of hard disks inside the tube, such as a dependence of the current with respect to the chemical potential of the reservoir, deviations from local equilibrium behaviors, the particle density and the local pressure of the systems.

The outline of this paper is as follows. 
In Sec. \ref{StochasticlBoundaryConditions}, PISB conditions for interactions with a particle reservoir are introduced in terms of the properties of particles injected from the reservoir. 
In Sec. \ref{EquilibriumPropertiesHardDiskSystems}, the PISB approach is checked by comparing various equilibrium statistical-mechanical quantities calculated by using the PISB conditions in equilibrium ideal gases and hard-disk systems, with those calculated analytically based on their corresponding grand canonical distributions.  
In Sec. \ref{ApplicationsParticleCurrents}, we apply the PISB method to nonequilibrium systems which consist of hard disks with a steady particle current inside a tube connected to a particle reservoir. 
In Sec. \ref{ConclusionsRemarks}, we give conclusions and remarks, and also add discussions on other approaches to grand canonical ensemble simulations and nonequilibrium open systems in relation to the PISB approach.

\section{Stochastic boundary conditions for systems coupled with a particle reservoir}
\label{StochasticlBoundaryConditions} 

\subsection{The momentum distribution of particles injected from a particle reservoir} 

We consider classical particle systems coupled to a particle reservoir with a temperature $T$ and a chemical potential $\mu$. 
Here, the coupling of such a system to a particle reservoir allows particles to come from and to go to the reservoir via the boundary $\ca{S}$ between the system and the reservoir. 
For simplicity, we assume that the boundary $\ca{S}$ is flat with a constant unit vector $\bs{n}$ on $\ca{S}$, in which $\bs{n}$ is perpendicular to $\ca{S}$ and is directed toward the system.  
Since particles inside the reservoir are in an equilibrium state with the temperature $T$, the momentum probability density function $f(\bs{p})$ of injected particles from the reservoir on the boundary $\ca{S}$ is given by \cite{CT80,TC82,MK84,MK87,BL91}    
\begin{eqnarray}
   f(\bs{p}) &=& \frac{\bs{n}\cdot\bs{p}}{\left(2\pi\right)^{(d-1)/2}\left(mk_{B}T\right)^{(d+1)/2}} 
   \exp\left(-\frac{\left|\bs{p}\right|^{2}}{2mk_{B}T}\right) 
   \nonumber\\
\label{InjecMomenDistr1}
\end{eqnarray}
with Boltzmann's constant $k_{B}$, the mass $m$ of each particle and the spatial dimension $d$ of the system. 
Here, since $\bs{p}$ is the momentum of a particle injected to the system, the inequality $\bs{n}\cdot\bs{p} > 0$ must be satisfied, i.e. $\bs{n}\cdot\bs{p} \in (0,+\infty)$, while other components of the momentum $\bs{p}$ except for $\bs{n}\cdot\bs{p}$, if any, can take any real value in $(-\infty,+\infty)$. 
Equation (\ref{InjecMomenDistr1}) comes from a count of the number of particles reaching the boundary $\ca{S}$ per unit time from the particle reservoir whose momentum distribution function is given by the Maxwell momentum distribution function \cite{K61}. 
It may also be noted that the momentum probability density function $f(\bs{p})$ of injected particles on the boundary $\ca{S}$ is not the Maxwell momentum distribution function itself \cite{TT98}.

\subsection{Time interval distribution of particle injections from a particle reservoir} 

In order to describe how particles are injected into the system from the particle reservoir, we also have to specify how frequently such injected particles reach the boundary $\ca{S}$ between the system and the reservoir. 
We assume that successive injections of particles are uncorrelated, and the probability density function $g(\tau)$ of time intervals $\tau$ ($\geq 0$) between two successive injections of particles via the boundary $\ca{S}$ is of the form   
\begin{eqnarray}
   g(\tau) = \nu e^{-\nu \tau} 
\label{TimeInterDistr1}
\end{eqnarray}
with a constant frequency $\nu$ as a positive coupling strength between the system and the particle reservoir. 
The exponential probability density function like in Eq. (\ref{TimeInterDistr1}) is also discussed in dynamics with couplings to a thermal reservoir in Refs. \cite{FS02,A80,PB93}.

By using Eqs. (\ref{InjecMomenDistr1}) and (\ref{TimeInterDistr1}) for systems with the spatial dimension $d\geq 2$, we estimate the average momentum injection $\ca{P}_{+}$ per unit area of the boundary $\ca{S}$ with the area $S$ per unit time as
\begin{eqnarray}
   \ca{P}_{+} = \frac{\overline{ \bs{n}\cdot\bs{p} }}{\overline{\tau}S} = \frac{\nu}{S}\sqrt{\frac{\pi m k_{B}T}{2}}
\label{MomenTrans1}
\end{eqnarray}
with $\overline{ \bs{n}\cdot\bs{p} } \equiv \int d\bs{p}\; \bs{n}\cdot\bs{p} f(\bs{p}) = \sqrt{\pi m k_{B}T/2}$ and $\overline{\tau} \equiv \int d\tau \; \tau g(\tau) = 1/\nu$. 
By introducing the average outgoing momentum $\ca{P}_{-}$ from the system per unit area of the boundary $\ca{S}$ per unit time, the pressure $P$ on the boundary $\ca{S}$ is given by the magnitude of the total momentum transfer $\ca{P}_{+} - \ca{P}_{-}$, i.e., $P = |\ca{P}_{+} - \ca{P}_{-}|$. 
Using the fact that at equilibrium states the average outgoing momentum $\ca{P}_{-}$ should be given by $-\ca{P}_{+}$ as a balance between the injecting momenta and outgoing momenta via the boundary $\ca{S}$, the pressure $P=P_{eq}$ at an equilibrium state should be given by
\begin{eqnarray}
   P_{eq} = 2\ca{P}_{+} .  
\label{Press1}
\end{eqnarray}
Equations (\ref{MomenTrans1}) and (\ref{Press1}) lead to 
\begin{eqnarray}
   \nu = \frac{P_{eq}(T,\mu)S}{\sqrt{2\pi m k_{B}T}} 
\label{NuPress1}
\end{eqnarray}
as the relation among the system-reservoir coupling strength $\nu$, the equilibrium pressure $P_{eq}$, and the temperature $T$.  
Here, we note that the equilibrium pressure $P_{eq}$ can be represented as the pressure of the particle reservoir as a function of the temperature $T$ and the chemical potential $\mu$: $P_{eq} = P_{eq}(T,\mu)$, based on the Gibbs-Duhem relation, so Eq. (\ref{NuPress1}) is regarded as an expression of the coupling strength $\nu$ as a function of $T$ and $\mu$.

\section{Equilibrium properties of systems with the PISB conditions}
\label{EquilibriumPropertiesHardDiskSystems} 
\subsection{Ideal gases}

As the first application of the PISB method discussed in Sec. \ref{StochasticlBoundaryConditions}, we consider two-dimensional ideal gases without any potential energy inside the systems. 
This type of systems is chosen, not only because of its simplicity, but also to show that the thermalization can occur by the PISB conditions themselves rather than by complex (chaotic) dynamics caused by particle-particle interactions, etc.

\subsubsection{Equilibrium properties of ideal gases based on the grand canonical distribution}
\label{EquilibriumPropertiesIdealGases}

By the equilibrium statistical mechanics with the grand canonical distribution, the grand potential $J$ of the two-dimensional ideal gases in an equilibrium state is given by 
\begin{eqnarray}
   J = -\frac{2\pi m  V e^{\beta\mu}}{\beta^{2} h^{2}} 
\label{GrandPotenIdeal1}
\end{eqnarray}
with the spatial area $V$ for particles to exist, the inverse temperature $\beta \equiv 1/(k_{B}T)$ and Planck's constant $h$. 
[A brief derivation of Eq.  (\ref{GrandPotenIdeal1}) from the corresponding grand canonical distribution is shown in Appendix \ref{IdelaGasGrand}.]
By the grand potential (\ref{GrandPotenIdeal1}) we obtain the pressure $P = - \partial J/\partial V|_{T,\mu}$ and the average $N = -\partial J/\partial \mu|_{T,V}$ of the number of particles as
\begin{eqnarray}
   P &=& \frac{2\pi m e^{\beta\mu}}{\beta^{2} h^{2}} , 
\label{PressIdeal1} \\ 
   N &=& \frac{2\pi m V e^{\beta\mu}}{\beta h^{2}} , 
\label{PartiNumbeIdeal1}
\end{eqnarray}
respectively. 
By Eqs. (\ref{PressIdeal1}) and (\ref{PartiNumbeIdeal1}) we obtain the equation of state $PV = Nk_{B}T$ for the ideal gases.

The grand potential $J = J(T,V,\mu)$ as a function of $T$, $V$ and $\mu$ also leads to the equilibrium generation function $\tilde{G}(z)$ of the number $\ca{N}$ of particles as a function of $z$, which is defined by the equilibrium average of $z^{\ca{N}}$, as
\begin{eqnarray}
   \tilde{G}(z) &=& \exp\left\{ -\beta \left[J\!\left(T, V,\mu + \frac{\ln z}{\beta}\right) 
    - J(T, V,\mu)\right]\right\} . 
    \nonumber \\
\label{GenerPartiNumbe2}
\end{eqnarray}
as shown in Appendix \ref{AnalyticalEquilibriumProperties}. 
Inserting Eq. (\ref{GrandPotenIdeal1}) into Eq. (\ref{GenerPartiNumbe2}) we obtain   
\begin{eqnarray}
   \tilde{G}(z) = e^{\lambda\left(z-1\right)} 
\label{GenerPartiIdeal2}
\end{eqnarray}
with $\lambda$ defined by
\begin{eqnarray}
	\lambda \equiv \frac{2\pi m V e^{\beta\mu}}{\beta h^{2} }. 
\label{Lambda2d1}
\end{eqnarray}
The generation function (\ref{GenerPartiIdeal2}) leads to the probability distribution $G(\ca{N})$ of the number $\ca{N}$ of particles as the Poisson distribution
\begin{eqnarray}
   G(\ca{N}) = \frac{1}{\ca{N}!}\left.\frac{d^{\ca{N}}\tilde{G}(z)}{dz^\ca{N}}\right|_{z=0}  = \frac{\lambda^{\ca{N}}}{\ca{N}!}e^{-\lambda}
\label{PartiNumbeDistrIdeal1}
\end{eqnarray}
for the two-dimensional ideal gases.  
(Another derivation of Eq. (\ref{PartiNumbeDistrIdeal1}) is shown in Appendix \ref{IdelaGasGrand}.)
From this distribution we can also derive the average $N = \lambda$ of the number $\ca{N}$ of particles, i.e., Eq. (\ref{PartiNumbeIdeal1}).

By inserting Eq. (\ref{PressIdeal1}) into Eq. (\ref{NuPress1}) we obtain the system-reservoir coupling strength $\nu$ for the ideal gases as  
\begin{eqnarray}
   \nu 
   = \frac{S e^{\beta\mu}\sqrt{2\pi m}}{\beta^{3/2}h^{2}} 
\label{NuPress2}
\end{eqnarray}
concretely.  
Particle injections from a particle reservoir in cases without particle-particle interactions are also discussed in Refs.  \cite{D95,I97,LJ13,CM08,HP09}.

\subsubsection{Ideal gases with the PISB conditions}

In order to apply the PISB method to the ideal gases concretely, we further assume that the two-dimensional ideal gases consist of point particles inside a square area with the side length $L$, so $V = L^{2}$ and $S=L$. 
We choose a single side of this square as the boundary $\ca{S}$ with the length $L$, and its other three sides are simply hard walls with which particles collide elastically. 
Particles are injected into the square area via the boundary $\ca{S}$ with the probability density function  (\ref{TimeInterDistr1}) of the time intervals $\tau$ between two successive injections of particles with the frequency (\ref{NuPress2}), and with the momentum distribution (\ref{InjecMomenDistr1}).  
Positions of such injected particles are chosen randomly and uniformly on the boundary $\ca{S}$.  
Any particle which reaches the boundary $\ca{S}$ from the square area is removed immediately from the systems.  
Here and hereafter, for two-dimensional models with a flat boundary $\ca{S}$ we take the $x$-axis as that perpendicular to the boundary $\ca{S}$, and the $y$-axis along $\ca{S}$.

For such square systems consisting of point particles, it would be meaningful to show that the relation (\ref{NuPress2}) among the coupling strength $\nu$, the temperature $T$ and the chemical potential $\mu$ can also be derived from another argument. 
We note that for the square systems consisting of point particles the average $\tilde{\tau}$ of the time interval for a particle injected from the reservoir at a point of the boundary $\ca{S}$ to return back to $\ca{S}$ can be estimated by the average of $2L/(\bs{n}\cdot\bs{p}/m)$, i.e., $\tilde{\tau} =  \int d\bs{p}\; [2mL/(\bs{n}\cdot\bs{p})] f(\bs{p}) = L \sqrt{2\pi m\beta}$. 
With this time interval $\tilde{\tau}$, the average $N$ of the number of particles inside the square systems could be estimated as the time interval $\tilde{\tau}$ divided by the average time interval $\overline{\tau} = 1/\nu$ between two successive injections of particles on $\ca{S}$, namely
\begin{eqnarray}
   N \approx \frac{\tilde{\tau}}{\overline{\tau}} = L \nu \sqrt{2\pi m\beta} . 
\label{PartiNumbeIdeal2}
\end{eqnarray}
By Eqs. (\ref{PartiNumbeIdeal1}), (\ref{PartiNumbeIdeal2}) and $V/L = S$ we obtain Eq. (\ref{NuPress2}) again.

Now, we discuss results of numerical calculations of equilibrium quantities of the ideal gases with the PISB conditions, and compare them with the corresponding analytical expressions given in Sec. \ref{EquilibriumPropertiesIdealGases}. 
For such numerical calculations, we chose the parameter values as  $m = 1$, $h=1$, $k_{B}T = 1$ and $L = 100$, and used the data over the time interval $\ca{T} = 10^{8}$ to calculate various time-average quantities and distributions. 
We took this time interval $\ca{T}$ after calculations of particle orbits at an early stage, so that in the time-interval $\ca{T}$ these quantities and distributions can be regarded to be stationary in time, supposing that the systems would be in equilibrium states.

\begin{figure}[!t]
\vspfigA
\begin{center}
\includegraphics[width=\widthfig]{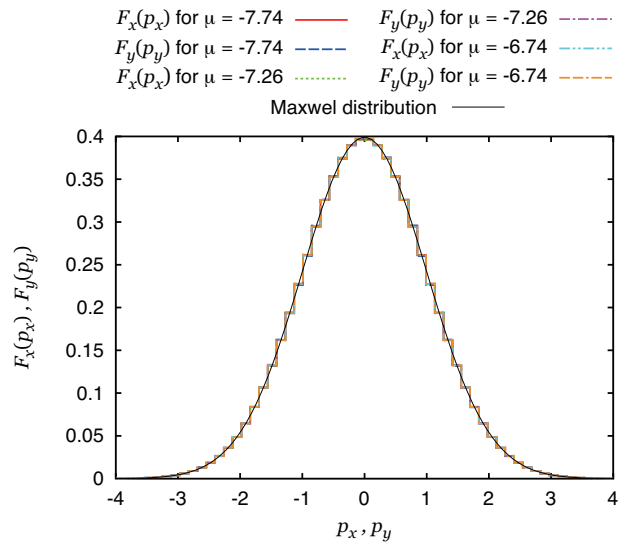}
\vspfigB
\caption{(Color online)
The probability density functions $F_{x}(p_{x})$ and $F_{y}(p_{y})$ of the components $p_{x}$ and $p_{y}$ of the particle momentum $(p_{x},p_{y})$, respectively, in ideal gases with the PISB conditions, for the chemical potentials $\mu = -7.74$, $-7.26$ and $-6.74$. 
All these probability density functions of the momentum components are almost indistinguishable with each other. 
The black thin line is the corresponding Maxwell momentum distribution. 
Here, and in all figures hereafter, we use dimensionless units with $m = 1$, $h=1$ and $k_{B}T = 1$. 
}
\label{Fig1MomenDistrIdeal}
\end{center}
\vspfigC
\end{figure}
%
In Fig. \ref{Fig1MomenDistrIdeal} we show the probability density functions $F_{x}(p_{x})$ and $F_{y}(p_{y})$ of the components $p_{x}$ and $p_{y}$ of the particle momentum $(p_{x},p_{y})$, respectively, for the ideal gases with the PISB conditions, for the chemical potentials $\mu = -7.74$, $-7.26$ and $-6.74$. 
(Here, the steps in these graphs of $F_{x}(p_{x})$ and $F_{y}(p_{y})$ come from the momentum width in which probabilities of the momentum components are calculated discretely. 
We use this type of presentations for momentum distributions in this paper.)  
The black thin line in Fig. \ref{Fig1MomenDistrIdeal} is the corresponding Maxwell momentum distribution. 
In this figure, all these probability density functions of the momentum components are almost indistinguishable with each other, and the probability density functions $F_{x}(p_{x})$ and $F_{y}(p_{y})$ almost coincide with the Maxwell momentum distribution 
\begin{eqnarray}
   F_{x}(p) = F_{y}(p) = \frac{1}{\sqrt{2\pi m k_{B}T}}\exp\left(-\frac{p^{2}}{2mk_{B}T}\right), 
\label{MaxweDistr1}
\end{eqnarray}
which is derived from the corresponding grand canonical distribution and is independent of $\mu$. 

Concerning the fact that the Maxwell momentum distribution (\ref{MaxweDistr1}) for particles inside the system is generated from the momentum distribution (\ref{InjecMomenDistr1}) for injected particles on the boundary $\ca{S}$, one may notice that the probability density of the momentum component $p_{x}$ of particles injected from the reservoir on the boundary $\ca{S}$, given from Eq. (\ref{InjecMomenDistr1}), is small for small absolute values of $p_{x}$, but injected particles with such small absolute values of $p_{x}$ stay for a long time inside the square area, so that the probability density $F_{x}(p_{x})$ of the momentum component $p_{x}$ of particles is large for small absolute values of $p_{x}$. 
Since the number of particles injected from the reservoir on the boundary $\ca{S}$ with small absolute values of $p_{x}$ are small relatively, it could take a long time for the Maxwell momentum distribution (\ref{MaxweDistr1}) for $F_{x}(p_{x})$ to be realized for particles inside the square area, for some initial conditions such as the one with no particle initially.

\begin{figure}[!t]
\vspfigA
\begin{center}
\includegraphics[width=\widthfig]{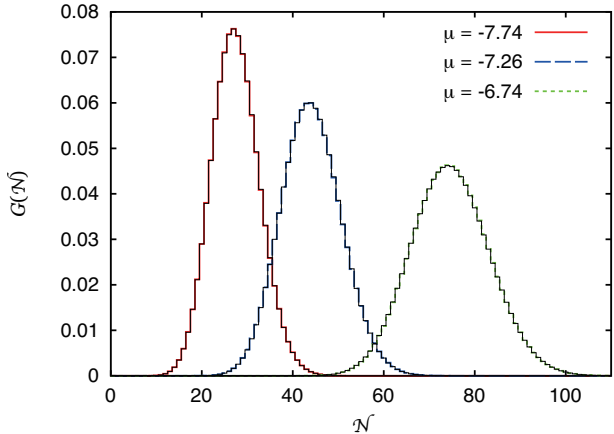}
\vspfigB
\caption{(Color online)
The probabilities $G(\ca{N})$ of the number $\ca{N}$ of particles in ideal gases with the PISB conditions, for the chemical potentials $\mu = -7.74$ (red solid line), $-7.26$ (blue dashed line) and $-6.74$ (green dotted line). 
The Poisson distribution (\ref{PartiNumbeDistrIdeal1}) is also shown as the black thin line in each chemical potential case, which is almost indistinguishable from the corresponding numerical results. 
}
\label{Fig2NumbeDistrIdeal}
\end{center}
\vspfigC
\end{figure}
%
In Fig. \ref{Fig2NumbeDistrIdeal} we show the probabilities $G(\ca{N})$ of the numbers $\ca{N}$ of particles in the systems with the PISB conditions for the chemical potentials $\mu = -7.74$ (red solid line), $-7.26$ (blue dashed line) and $-6.74$ (green dotted line). 
The corresponding graphs of Eq. (\ref{PartiNumbeDistrIdeal1}), given from the grand canonical distribution of the ideal gases, are also shown in this figure as the black thin lines, which show an excellent agreement with the ones calculated numerically by using the PISB approach. 
This also suggests that Eq. (\ref{PartiNumbeIdeal1}) derived from Eq. (\ref{PartiNumbeDistrIdeal1}) is satisfied for the ideal gases with the PISB conditions. 

\begin{figure}[!t]
\vspfigA
\begin{center}
\includegraphics[width=\widthfig]{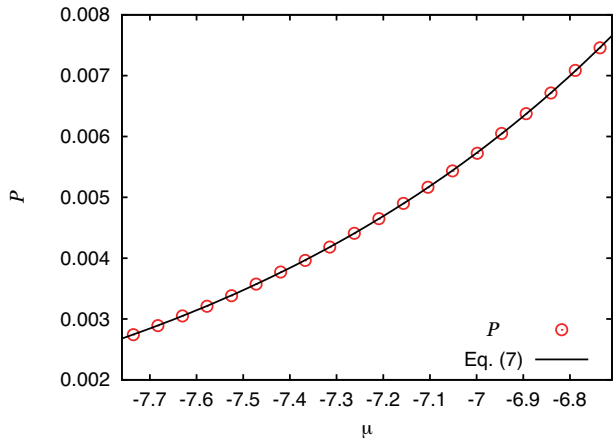}
\vspfigB
\caption{(Color online)
The pressure $P$ of ideal gases with the PISB conditions as a function of the chemical potential $\mu$ (red circles). 
The black line is the pressure (\ref{PressIdeal1}) based on the corresponding grand canonical distribution.  
}
\label{Fig3PressIdeal}
\end{center}
\vspfigC
\end{figure}
%
The pressure of the ideal gases can be calculated numerically as the absolute value of the time-average impulse which particles give to a wall by their collisions per unit time and per unit length of the wall. 
As such a wall for actual numerical calculations of the pressure, we chose a wall perpendicular to the boundary $\ca{S}$. 
This wall was chosen to analyze the pressure in the direction perpendicular to the one for the pressure from which we introduced the system-reservoir coupling strength $\nu$ as Eq. (\ref{NuPress1}). 
In Fig. \ref{Fig3PressIdeal} we show such a pressure $P$ of the ideal gases with the PISB conditions as a function of the chemical potential $\mu$ (red circles). 
Here, the pressure (\ref{PressIdeal1}) calculated analytically by the corresponding grand canonical distribution is also shown as the black line, and its values are almost indistinguishable from the ones calculated numerically by the PISB approach. 
The results shown in Figs. \ref{Fig2NumbeDistrIdeal} and \ref{Fig3PressIdeal} also suggest that the ideal gases in equilibrium states with the PISB conditions satisfy the equation of state $PV = Nk_{B}T$.

\subsection{Hard-disk systems}
\label{HardDiskSystems}

Now we discuss applications of the PISB method to systems consisting of hard disks, as an example of systems with particle-particle interactions. 
For the hard-disk systems discussed in this paper, we assume that there is not a potential energy other than the one corresponding to the particle-particle interactions inside the systems.    

\subsubsection{Equilibrium properties of hard-disk systems based on the grand canonical distribution} 

In this paper, we use the grand potential $J$ of the two-dimensional equilibrium systems consisting of identical hard disks with the radius $r$ as 
\begin{eqnarray}
   J =  - \frac{2\pi mV}{\beta^{2} h^{2}}\; e^{\beta\mu}
       \left(1- \frac{4\pi^{2} r^{2} m}{\beta h^{2}} e^{\beta\mu}\right)
\label{GrandPotenDisk1}
\end{eqnarray} 
in low-density cases, as a function of the temperature $T$ $[=1/(k_{B}\beta)]$, the spatial area $V$ and the chemical potential $\mu$. 
Note that this grand potential $J$ of hard-disk systems, which is proportional to $V$, has an extensive property. 
A derivation of Eq. (\ref{GrandPotenDisk1}) and specific conditions for it are discussed in Appendix \ref{HardDiskGrand}, based on  the corresponding grand canonical distribution. 
The grand potential (\ref{GrandPotenDisk1}) leads to the pressure $P= - \partial J/\partial V|_{T,\mu}$, the average $N = -\partial J/\partial \mu|_{T,V}$ of numbers of particles, and the variance $\langle(\ca{N}-N)^{2}\rangle =  - \beta^{-1}\partial^{2} J/\partial \mu^{2}|_{T,V}$ of numbers of particles as
\begin{eqnarray}
   P = \frac{2\pi m}{\beta^{2} h^{2}}\; e^{\beta\mu}
       \left(1- \frac{4\pi^{2}r^{2} m}{\beta h^{2}} e^{\beta\mu}\right),
      \label{PressDisk1} 
\end{eqnarray}
\begin{eqnarray}
   N =  \frac{2\pi m V}{\beta h^{2}} \; e^{\beta\mu}
       \left(1-\frac{8\pi^{2}r^{2} m}{\beta h^{2}} e^{\beta\mu}\right),
      \label{NumDisk1}  
\end{eqnarray}
\begin{eqnarray}
   \langle(\ca{N}-N)^{2}\rangle 
      =  \frac{2\pi m V}{\beta h^{2}}\; e^{\beta\mu}
      \left(1-\frac{16\pi^{2}r^{2} m }{\beta h^{2}} e^{\beta\mu}\right),
      \label{VariaDisk1}  
\end{eqnarray}
respectively. 
By Eqs. (\ref{PressDisk1}) and (\ref{NumDisk1}) we obtain the relation 
\begin{eqnarray}
   \frac{N k_{B}T}{PV} \approx 1-2\pi r^{2}\frac{N}{V}
\label{VanDerWaals1}
\end{eqnarray}
approximately, neglecting the terms including the order of $r^{4}$, leading van der Waal's equation of state as $P(V-2\pi r^{2}N) \approx Nk_{B}T$ for two-dimensional imperfect gases without any attractive particle-particle interaction.

By Eqs. (\ref{NuPress1}) and  (\ref{PressDisk1}) the system-reservoir coupling strength $\nu$ for the hard-disk systems is given by 
\begin{eqnarray}
   \nu = \frac{S e^{\beta\mu}\sqrt{2\pi m}}{\beta^{3/2}h^{2}} \left(1- \frac{4\pi^{2}r^{2} m}{\beta h^{2}} e^{\beta\mu}\right)
\label{NuPress3}
\end{eqnarray}
concretely. 
Eq. (\ref{NuPress3}) means that the frequency $\nu$ in the probability density function (\ref{TimeInterDistr1}) of time intervals $\tau$ between two successive injections of particles decreases as the particle size $\pi r^{2}$ increases.

\subsubsection{Hard-disk systems with the PISB conditions}
\label{HardDiskStochasticlBoundaryConditions}

Now, we discuss statistical-mechanical properties of hard-disk systems with the PISB conditions. 
We consider hard disks with the radius $r$ inside a rectangular area with the side lengths $L_{x}$ and $L_{y}$, and one of four sides of the rectangular area with the side length $L_{y}$ becomes the boundary $\ca{S}$ for disks to be injected from a particle reservoir with the temperature $T$ and the chemical potential $\mu$, while the other three sides are hard walls. 
For such a system, we introduce the length $S$ of the boundary $\ca{S}$ and the spatial area $V$ of the rectangular area as them for the centers of particles to exist, so that they are given by $S=L_{y}-2r$ and $V = S(L_{x}-r) = (L_{x}-r)(L_{y}-2r)$ since the centers of particles cannot be closer to their hard walls than the disk radius $r$. 
The centers of injected hard disks with the momentum probability density function (\ref{InjecMomenDistr1}) and the time-interval probability density function (\ref{TimeInterDistr1}) appear randomly on the boundary $\ca{S}$ as far as such injected hard disks are not overlapped with other hard disks inside the system \cite{MemoA}. 
Any hard disk whose center reaches the boundary $\ca{S}$ from the inside of the system is removed immediately from the system.

We investigated numerically equilibrium quantities and distributions of the hard-disk systems with $r=0.5$, $m=1$, $h=1$, $k_{B}T=1$, $L_{x}=100+r$ and $L_{y}=100+2r$ (so $S=100$ and $V=S^{2}$) for various values of the chemical potential $\mu$. 
For these numerical calculations we used the data over the time-interval $\ca{T}=10^{8}$ in which these quantities and distributions would be stationary in time. 
Our numerical results show that the probability density functions $F_{x}(p_{x})$ and $F_{y}(p_{y})$ of the components $p_{x}$ and $p_{y}$ of the particle momentum $(p_{x},p_{y})$ of the hard-disk systems, respectively, satisfy the Maxwell momentum distributions (\ref{MaxweDistr1}). 
(A figure to show this point for the hard-disk systems is quite similar to Fig. \ref{Fig1MomenDistrIdeal} for ideal gases, so we omit to show such a figure explicitly in this paper.)

\begin{figure}[!t]
\vspfigA
\begin{center}
\includegraphics[width=\widthfig]{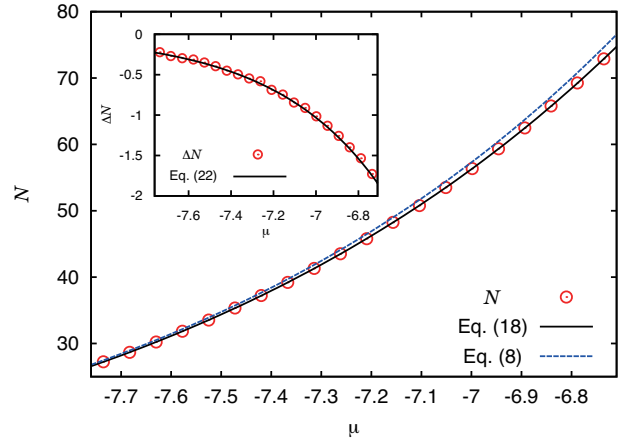}
\vspfigB
\caption{(Color online) 
The average number $N$ of disks (red circles) in hard-disk systems with the PISB conditions, as well as the graph of Eq. (\ref{NumDisk1}) (black solid line), as a function of the chemical potential $\mu$. 
Here, the graph of Eq. (\ref{PartiNumbeIdeal1}) for the corresponding ideal gases is also shown by the blue dashed line. 
The inset: The deviation $\Delta N \equiv N - N|_{r=0}$ (red circles) of the average number $N$ from its ideal-gas case, as well as the graph of Eq. (\ref{NumDiskDevia1}) (black solid line), as functions of $\mu$.}
\label{Fig4NumbeAveraDisks}
\end{center}
\vspfigC
\end{figure}
%
The main figure in Fig. \ref{Fig4NumbeAveraDisks} is the average number $N$ of disks (red circles) in the hard-disk systems with the PISB conditions, as well as the graph of Eq. (\ref{NumDisk1}) (black solid line), as a function of the chemical potential $\mu$. 
Here, for a comparison, we also show the graph of Eq. (\ref{PartiNumbeIdeal1}) for the corresponding ideal gases by the blue dashed line. 
In low-density cases, a difference between Eqs. (\ref{PartiNumbeIdeal1})  and  (\ref{NumDisk1}) is small because of weak effects of disk-disk collisions, so in the inset of Fig. \ref{Fig4NumbeAveraDisks} as a quantity to show such collision effects explicitly we plot the deviation $\Delta N \equiv N - N|_{r=0}$ (red circles) of the average number $N$ from the one of the corresponding ideal gas given by $N|_{r=0} \equiv 2\pi m V e^{\beta\mu}/(\beta h^{2})$, i.e., Eq. (\ref{PartiNumbeIdeal1}), as well as the graph of 
\begin{eqnarray}
   \Delta N 
   = -\frac{16\pi^{3}r^{2} m^{2}V }{\beta^{2} h^{4}}e^{2\beta\mu}
\label{NumDiskDevia1}
\end{eqnarray}
given from Eq. (\ref{NumDisk1}) (black solid line), as a function of $\mu$. 
Figure \ref{Fig4NumbeAveraDisks} shows that the average number $N$ of disks in the hard-disk systems with the PISB conditions almost coincide with Eq. (\ref{NumDisk1}) based on the corresponding grand canonical distribution for $\mu \in (-7.74,-6.74)$. 
%

\begin{figure}[!t]
\vspfigA
\begin{center}
\includegraphics[width=\widthfig]{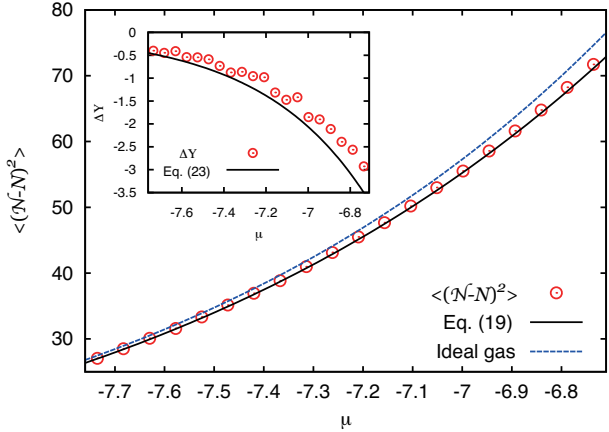}
\vspfigB
\caption{(Color online) 
The variance $\langle(\ca{N}-N)^{2}\rangle$ of numbers of particles (red circles) as a function of the chemical potential $\mu$ in hard-disk systems with the PISB conditions. 
Here, the graph of Eq. (\ref{VariaDisk1}) (black solid line), as well as the graph of the variance of numbers of particles in the corresponding ideal-gas cases (blue dashed line), given from the corresponding grand canonical distributions, are also shown. 
The inset: The deviation $\Delta \Upsilon \equiv \langle(\ca{N}-N)^{2}\rangle - \langle(\ca{N}-N)^{2}\rangle|_{r=0}$ (red circles) and the graph of Eq. (\ref{VariaDiskDevia1}) (black solid line) as functions of $\mu$. 
}
\label{Fig5NumbeVariaDisks}
\end{center}
\vspfigC
\end{figure}
%
In the main figure of Fig. \ref{Fig5NumbeVariaDisks} we show the variance $\langle(\ca{N}-N)^{2}\rangle$ of numbers of particles (red circles) as a function of the chemical potential $\mu$ in the hard-disk systems with the PISB conditions.   
Here, the graph of Eq. (\ref{VariaDisk1}) (black solid line), as well as the graph of the variance of numbers of particles in the corresponding ideal-gas cases (blue dashed line), both of which are given by the corresponding grand canonical distributions, are also shown. 
The inset of this figure is the deviation $\Delta \Upsilon \equiv \langle(\ca{N}-N)^{2}\rangle - \langle(\ca{N}-N)^{2}\rangle|_{r=0}$ of the variance of numbers of particles (red circles) from the one of the corresponding ideal gas given by $\langle(\ca{N}-N)^{2}\rangle|_{r=0} \equiv 2\pi m V e^{\beta\mu}/(\beta h^{2})$, as well as the graph of 
\begin{eqnarray}
   \Delta \Upsilon  
      =  -\frac{32\pi^{3}r^{2} m^{2}V }{\beta^{2} h^{4}}e^{2\beta\mu} = 2 \Delta N
      \label{VariaDiskDevia1}  
\end{eqnarray}
given from Eq. (\ref{VariaDisk1}) (black solid line), as functions of $\mu$. 
Figure \ref{Fig5NumbeVariaDisks} shows that Eq. (\ref{VariaDisk1}) based on the corresponding grand canonical distribution gives a good approximation of the variance $\langle(\ca{N}-N)^{2}\rangle$ of numbers of particles for the hard-disk systems with the PISB conditions, although it slightly underestimates the values of the quantity $\Delta \Upsilon$. 
Actually, the absolute values of all relative errors of the numerical values of $\langle(\ca{N}-N)^{2}\rangle$ in Fig. \ref{Fig5NumbeVariaDisks} from Eq. (\ref{VariaDisk1}) are less than 1 percent.

\begin{figure}[!t]
\vspfigA
\begin{center}
\includegraphics[width=\widthfig]{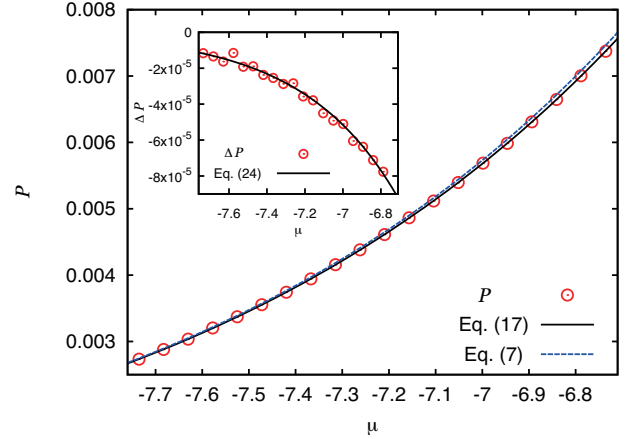}
\vspfigB
\caption{(Color online)
The pressure $P$ of hard-disk systems  with the PISB conditions (red circles), and the graphs of Eq. (\ref{PressDisk1}) (black solid line) and Eq. (\ref{PressIdeal1}) (blue dashed line), as functions of the chemical potential $\mu$. 
The inset: The difference $\Delta P \equiv P - P|_{r=0}$ between the pressures $P$ and $P|_{r=0}\equiv 2\pi m e^{\beta\mu}/(\beta^{2} h^{2})$ (red circles), and the graph of Eq. (\ref{PressDiskDeriv1}) (black solid line), as functions of $\mu$. 
}
\label{Fig6PressDisks}
\end{center}
\vspfigC
\end{figure}
%
We calculated the pressure $P$ of the hard-disk systems as the absolute value of the time-average impulse by particle collisions with a wall perpendicular to the boundary $\ca{S}$ per unit time and per unit length of the wall. 
Here, in order to calculate impulses per unit length of the wall, we divide total impulses by the effective side length $L_{x}-r$ for the centers of disks to exist in the $x$-direction, not by the actual side length $L_{x}$. 
The main figure of Fig. \ref{Fig6PressDisks} is such a pressure $P$ (red circles) of the hard-disk systems with the PISB conditions, as well as Eq. (\ref{PressDisk1}) (black solid line) and Eq. (\ref{PressIdeal1}) (blue dashed line) given from the corresponding grand canonical distributions, as functions of the chemical potential $\mu$. 
The difference $\Delta P \equiv P - P|_{r=0}$ between the pressures $P$ and $P|_{r=0}$ for a hard-disk system and the corresponding ideal gas, respectively, is small for the range of the chemical potential $\mu$ presented in the main figure of Fig. \ref{Fig6PressDisks}, so we plotted this difference $\Delta P$ itself with $P|_{r=0} \equiv 2\pi m e^{\beta\mu}/(\beta^{2} h^{2})$, i.e. Eq. (\ref{PressIdeal1}), in the inset of Fig. \ref{Fig6PressDisks}, for the hard-disk systems with the PISB conditions (red circles), as well as the graph of 
\begin{eqnarray}
   \Delta P = -\frac{8\pi^{3}r^{2}m^{2}}{\beta^{3} h^{4}} e^{2\beta\mu} 
\label{PressDiskDeriv1}
\end{eqnarray}
given from Eq. (\ref{PressDisk1}) (black solid line). 
These graphs show that the pressure (\ref{PressDisk1}) by the corresponding grand canonical distribution is reproduced extremely well by the one of the hard-disk systems with the PISB conditions discussed in Sec. \ref{StochasticlBoundaryConditions}.  
Our results in Figs. \ref{Fig4NumbeAveraDisks} and \ref{Fig6PressDisks} also indicate that the equation of state of the hard-disk systems with the PISB conditions is given by Eq. (\ref{VanDerWaals1}), i.e. van der Waal's equation of state, in a good approximation, although we omit a figure to show it explicitly in this paper.

\section{Particle currents from a particle reservoir} 
\label{ApplicationsParticleCurrents} 

In this section, we discuss an application of the PISB method to nonequilibrium phenomena with particle currents.

\begin{figure}[!t]
\vspfigA
\begin{center}
\includegraphics[width=\widthfig]{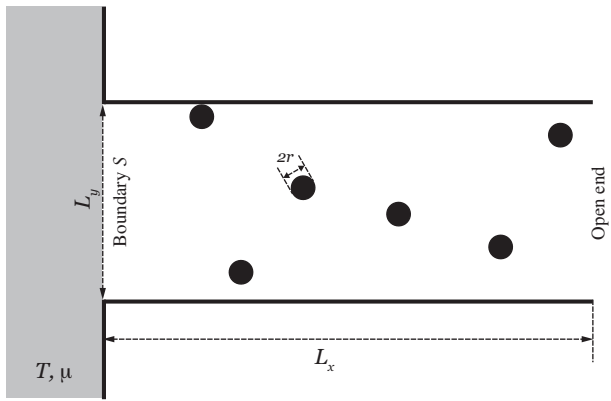}
\vspfigB
\caption{
Hard disks with the radius $r$ in an open-end tube with the length $L_{x}$ and the width $L_{y}$. 
The left end of the tube is connected to a particle reservoir (grey region) with the temperature $T$ and the chemical potential $\mu$ via the boundary $\ca{S}$, and the right end of the tube is open. 
Hard disks injected from the particle reservoir can leave not only from the left end, but also from the right end after moving inside the tube. 
}
\label{Fig7EscapeDisk}
\end{center}
\vspfigC
\end{figure}
%
\subsection{Currents caused by hard disks escaping from a particle reservoir} 

As simple nonequilibrium phenomena driven by a particle reservoir, we consider currents of particles escaping from a particle reservoir via a tube. 
The systems, which we consider in this section, consist of hard disks with the radius $r$ in an open-end two-dimensional tube whose length and width are $L_{x}$ and $L_{y}$, respectively. 
An end of this tube (the end $\ca{L}$) is connected to a particle reservoir with the temperature $T$ and the chemical potential $\mu$, and the comings and goings of hard disks at this end as the boundary $\ca{S}$ are described as these in Sec. \ref{HardDiskStochasticlBoundaryConditions} with the PISB conditions. 
Another end of this tube (the end $\ca{R}$) is open so that wherever the center of any hard disk reaches this end its hard disk are removed immediately from the systems, and there is no particle injection from this end into the systems. 
In such a system, some injected particles from the reservoir can leave the system from the end $\ca{R}$ of the tube, as escaping particles from the particle reservoir via the tube, and a nonequilibrium steady state with a nonzero particle current from the end $\ca{L}$ to the end $\ca{R}$ is realized after a long time. 
A schematic illustration of this type of systems is given in Fig. \ref{Fig7EscapeDisk}. 
For later uses, we introduce the $x$-axis directing from the end $\ca{L}$ to the end $\ca{R}$, and put the origin at a point in the end $\ca{L}$ of the tube.

In this section we analyze nonequilibrium statistical-mechanical properties of the systems numerically with the parameter values of $m = 1$, $h=1$, $r=0.5$, $L_{x}=1000$, $L_{y}=10+2r$ and $k_{B}T=1$. 
Here, we use a long tube with $L_{x}>\!>L_{y}$ so that many disk-disk collisions can occur inside the tube. 
Disk-disk collisions allow some disks injected from the reservoir to go back to the reservoir after moving inside the tube, so they have an effect to suppress the particle current from the end $\ca{L}$ to the end $\ca{R}$.

\begin{figure}[!t]
\vspfigA
\begin{center}
\includegraphics[width=\widthfig]{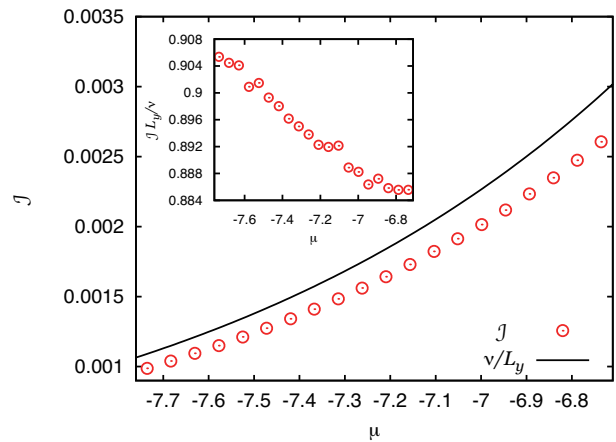}
\vspfigB
\caption{(Color online) 
The particle current density $\ca{J}$ (red circles) of hard disks in an open tube connected to a particle reservoir with the chemical potential $\mu$ and the temperature $T$, and the injected current density $\nu/L_{y}$ (black solid line) from the reservoir, as functions of $\mu$. 
The inset: the ratio $\ca{J}L_{y}/\nu$ between the particle current density $\ca{J}$ and the injected current density $\nu/L_{y}$ as a function of $\mu$. }
\label{Fig8Current}
\end{center}
\vspfigC
\end{figure}
%
In the main figure of Fig. \ref{Fig8Current} we show the particle current density $\ca{J}$ from the end $\ca{L}$ to the end $\ca{R}$ as a function of the chemical potential $\mu$.  
Here, we calculated numerically the current density $\ca{J}$ as $\ca{J} \approx (\ca{N}_{+} -\ca{N}_{-})/(\ca{T}L_{y})$ with the number $\ca{N}_{+}$ of injected disks from the particle reservoir and the number $\ca{N}_{-}$ of disks leaving for the particle reservoir from the end $\ca{L}$ of the tube over the time interval $\ca{T} = 10^{8}$ in which the systems could be regarded as being in steady states. 
In this figure, we also show the injected current density $\nu/L_{y} \approx \ca{N}_{+}/(\ca{T}L_{y})$ from the particle reservoir as a comparison, so that the difference $\ca{J}-\nu/L_{y}$ gives the particle current density leaving for the particle reservoir from the end $\ca{L}$ of the tube, caused by disk-disk collisions. 
Figure \ref{Fig8Current} shows that not only the injected current density $\nu/L_{y}$ given from Eq. (\ref{NuPress3}), but also the particle current density $\ca{J}$ itself is a monotonically increasing function of the chemical potential $\mu$ in the region of $\mu$ presented in this figure. 

To discuss an effect of particle-particle interactions we also plot the ratio $\ca{J}L_{y}/\nu$ (as the quantity taking the value $1$ for the corresponding ideal-gas systems) between the particle current density $\ca{J}$ and the injected current density $\nu/L_{y}$ as a function of the chemical potential $\mu$, in the inset of Fig. \ref{Fig8Current}.  
This inset suggests that the ratio $\ca{J}L_{y}/\nu$ is a monotonically decreasing function of $\mu$. 
This behavior of $\ca{J}L_{y}/\nu$ could be explained by the fact that an increase of the chemical potential $\mu$ makes the number of disks (and the number of disk-disk collisions per unit time) increase inside the tube, so that the probability for a disk injected from the particle reservoir to return back to the same reservoir also increases.

\subsection{Statistical-mechanical properties of particles in escaping currents from a particle reservoir} 
\label{StatisticalMechanicalPropertiesEscapingCurrents}

Now, we proceed to discuss statistical-mechanical properties of the systems consisting of hard disks escaping from a particle reservoir via a tube. 
Since the systems are in nonequilibrium states, various quantities of the systems depend on spatial positions. 
To discuss those position-depending properties of the systems, we introduce the region $W_{j}$ inside the tube as that satisfying the inequality $(j-1) \Delta L_{x} \leq x < j \Delta L_{x}$ with $\Delta L_{x} \equiv L_{x}/\eta$, and consider statistical-mechanical properties of hard disks in each of the evenly divided regions $W_{j}$ of the tube, for $j = 1,2,\cdots,\eta$ with $\eta=20$.

\begin{figure}[!t]
\vspfigA
\begin{center}
\includegraphics[width=\widthfig]{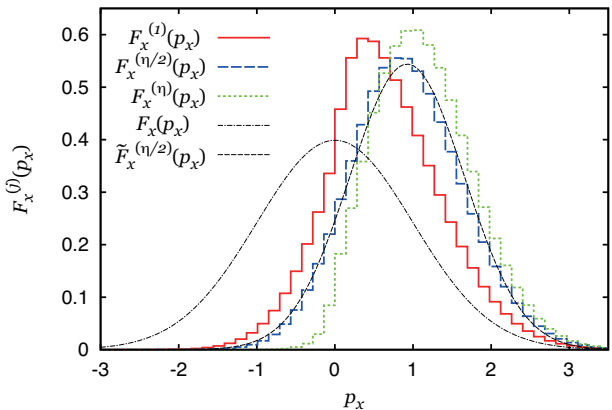}
\vspfigB
\caption{(Color online) 
The probability density functions $F_{x}^{(1)}(p_{x})$ (red solid line), $F_{x}^{(\eta/2)}(p_{x})$ (blue dashed line) and $F_{x}^{(\eta)}(p_{x})$ (green dotted line) of $x$-components of momentum vectors of hard disks in the regions $W_{1}$, $W_{\eta/2}$ and $W_{\eta}$, respectively, in the model with hard disks in an open tube connected to a particle reservoir with the chemical potential $\mu$ and the temperature $T$. 
The corresponding Maxwell momentum distribution $F_{x}(p_{x})$ with the temperature $T$, and the Gaussian distribution function $\tilde{F}_{x}^{(\eta/2)}(p_{x})$ with the average and the variance given from the data for $F_{x}^{(\eta/2)}(p_{x})$ are also shown as the black thin dashed-dotted line and the black thin dashed line, respectively. 
}
\label{Fig9MomenPxDistriNoneqil}
\end{center}
\vspfigC
\end{figure}
%
We introduce the probability distribution functions $F_{x}^{(j)}(p_{x})$ and $F_{y}^{(j)}(p_{x})$ of $x$-components and $y$-components of momentum vectors of hard disks in the region $W_{j}$, respectively, for $j=1,2,\cdots,\eta$, and discuss local momentum distributions of hard disks in the tube by these probability distribution functions in the case of $\mu = -6.74$. 
In Fig. \ref{Fig9MomenPxDistriNoneqil} we show the probability density functions $F_{x}^{(1)}(p_{x})$ (red solid line), $F_{x}^{(\eta/2)}(p_{x})$ (blue dashed line) and $F_{x}^{(\eta)}(p_{x})$ (green dotted line) for hard disks in the open tube connected to a particle reservoir with the chemical potential $\mu$ and the temperature $T$. 
For comparisons, we also plotted in this figure the Maxwell momentum distribution $F_{x}(p_{x})$ (black thin dashed-dotted line) given by Eq. (\ref{MaxweDistr1}) with the temperature $T$. 
We can easily see in Fig. \ref{Fig9MomenPxDistriNoneqil} that the peak positions of $F_{x}^{(j)}(p_{x})$, $j=1,\eta/2,\eta$ are sifted to positive values from the zero value for $F_{x}(p_{x})$ because of the escaping current of hard disks in the positive direction of the $x$ axis.

In Fig. \ref{Fig9MomenPxDistriNoneqil} we can also recognize asymmetric shapes in the peak region of $F_{x}^{(1)}(p_{x})$ and in the tail regions of $F_{x}^{(\eta)}(p_{x})$, meaning that the probability density function $F_{x}^{(1)}(p_{x})$ ($F_{x}^{(\eta)}(p_{x})$) in the end area $W_{1}$ ($W_{\eta}$) of the tube is not a Gaussian distribution. 
On the other hand, the probability density function $F_{x}^{(\eta/2)}(p_{x})$ in a middle region $W_{\eta/2}$ of the tube looks to be symmetric. 
To discuss this point further, we plotted in Fig. (\ref{Fig9MomenPxDistriNoneqil}) the Gaussian distribution $\tilde{F}_{x}^{(\eta/2)}(p_{x}) \equiv \sqrt{1/(2\pi\varsigma_{x}^{(\eta/2)})}\exp[- (p_{x}-\zeta_{x}^{(\eta/2)})^{2}/(2\varsigma_{x}^{(\eta/2)})]$  (black thin dashed line) with the average $\zeta_{x}^{(\eta/2)}$ and the variance $\varsigma_{x}^{(\eta/2)}$ of $p_{x}$ calculated numerically from the data giving the function $F_{x}^{(\eta/2)}(p_{x})$. 
The function $\tilde{F}_{x}^{(\eta)}(p_{x}) $ fits well to the function $F_{x}^{(\eta/2)}(p_{x})$ in Fig. \ref{Fig9MomenPxDistriNoneqil}, implying that the probability density function $F_{x}^{(\eta/2)}(p_{x})$ is well approximated by a Gaussian distribution.

\begin{figure}[!t]
\vspfigA
\begin{center}
\includegraphics[width=\widthfig]{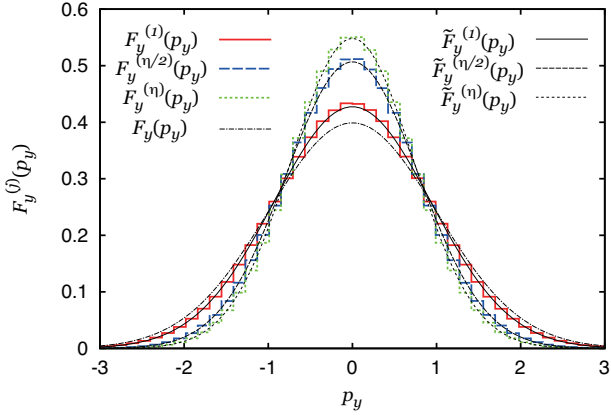}
\vspfigB
\caption{(Color online) 
The probability density functions $F_{y}^{(1)}(p_{y})$ (red solid line), $F_{y}^{(\eta/2)}(p_{y})$ (blue dashed line) and $F_{y}^{(\eta)}(p_{y})$ (green dotted line) of $y$-components of momentum vectors of hard disks in the regions $W_{1}$, $W_{\eta/2}$ and $W_{\eta}$, respectively, in the model with hard disks in an open tube connected to a particle reservoir with  the chemical potential $\mu$ and the temperature $T$. 
The corresponding Maxwell momentum distribution $F_{y}(p_{y})$ with the temperature $T$ is shown as the black thin dashed-dotted line. 
The Gaussian distribution functions $\tilde{F}_{y}^{(1)}(p_{x})$, $\tilde{F}_{y}^{(\eta/2)}(p_{x})$ and $\tilde{F}_{y}^{(\eta)}(p_{x})$ with the averages and the variances given from the data for $F_{x}^{(1)}(p_{x})$, $F_{x}^{(\eta/2)}(p_{x})$ and $F_{x}^{(\eta)}(p_{x})$ are also shown as the black thin solid line, the black thin dashed line and the black thin dotted line, respectively. 
}
\label{Fig10MomenPyDistriNonequil}
\end{center}
\vspfigC
\end{figure}
%
In Fig. \ref{Fig10MomenPyDistriNonequil} we show the graphs of the probability density functions $F_{y}^{(1)}(p_{y})$ (red solid line), $F_{y}^{(\eta/2)}(p_{y})$ (blue dashed line) and $F_{y}^{(\eta)}(p_{y})$ (green dotted line) of $y$-components of momentum vectors of hard disks in the regions $W_{1}$, $W_{\eta/2}$ and $W_{\eta}$, respectively, in the model with hard disks in an open tube connected to a particle reservoir.  
In this figure, the peak positions of the probability density functions $F_{y}^{(1)}(p_{y})$, $F_{y}^{(\eta/2)}(p_{y})$ and $F_{y}^{(\eta)}(p_{y})$ of $y$-components of momentum vectors of hard disks are at $p_{y} = 0$. 
Besides, the variances of $p_{y}$ in them are small for their positions far from the boundary $\ca{S}$ with the particle reservoir with the chemical potential $\mu$ and the temperature $T$.  
(For a comparison, the Maxwell momentum distribution $F_{y}(p_{y})$ with the temperature $T$ is also shown as the black thin dashed-dotted line in Fig. \ref{Fig10MomenPyDistriNonequil}. 
The variances of $p_{y}$ in the probability density functions $F_{y}^{(1)}(p_{y})$, $F_{y}^{(\eta/2)}(p_{y})$ and $F_{y}^{(\eta)}(p_{y})$ are smaller than that in $F_{y}(p_{y})$.) 
In Fig. \ref{Fig10MomenPyDistriNonequil} we also show the Gaussian distribution functions $\tilde{F}_{y}^{(1)}(p_{x})$, $\tilde{F}_{y}^{(\eta/2)}(p_{x})$ and $\tilde{F}_{y}^{(\eta)}(p_{x})$ with the averages and the variances given from the data for $F_{x}^{(1)}(p_{x})$, $F_{x}^{(\eta/2)}(p_{x})$ and $F_{x}^{(\eta)}(p_{x})$ as the black thin solid line, the black thin dashed line and the black thin dotted line, respectively. 
The functions $\tilde{F}_{y}^{(1)}(p_{x})$, $\tilde{F}_{y}^{(\eta/2)}(p_{x})$ and $\tilde{F}_{y}^{(\eta)}(p_{x})$ fit well to the corresponding probability density functions $F_{y}^{(1)}(p_{y})$, $F_{y}^{(\eta/2)}(p_{y})$ and $F_{y}^{(\eta)}(p_{y})$, respectively, suggesting that the probability density functions of $y$-components of momentum vectors of hard disks are well described by Gaussian distributions locally.

\begin{figure}[!t]
\vspfigA
\begin{center}
\includegraphics[width=\widthfigB]{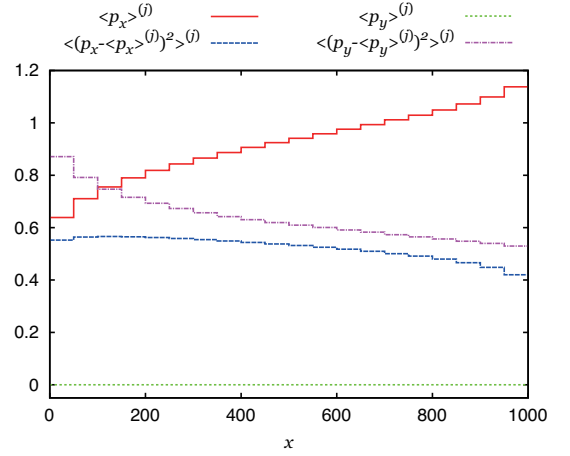}
\vspfigB
\caption{(Color online) 
The average $\langle p_{x} \rangle^{(j)}$ (red solid line) and the variance $\langle (p_{x} - \langle p_{x} \rangle^{(j)})^{2} \rangle^{(j)}$ (blue dashed line) of the momentum component $p_{x}$, and the average $\langle p_{y} \rangle^{(j)}$ (green dotted line) and the variance $\langle (p_{y} - \langle p_{y} \rangle^{(j)})^{2} \rangle^{(j)}$ (purple dashed-dotted line) of the momentum component $p_{y}$, in the region $W_{j}$ for $j=1,2,\cdots,\eta$ as functions of $x$, in the model with hard disks in an open tube connected to a particle reservoir. 
}
\label{Fig11PxPyAveFluNoneq}
\end{center}
\vspfigC
\end{figure}
%
Now, we discuss the average $\langle p_{x} \rangle^{(j)}$ and the variance $\langle (p_{x}- \langle p_{x} \rangle^{(j)})^{2} \rangle^{(j)}$ (the average $\langle p_{y} \rangle^{(j)}$ and the variance $\langle (p_{y}- \langle p_{y} \rangle^{(j)})^{2} \rangle^{(j)}$) of the momentum component $p_{x}$ (the momentum component $p_{y}$) of hard disks in the region $W_{j}$ for $j=1,2,\cdots,\eta$. 
In Fig.  \ref{Fig11PxPyAveFluNoneq} we show the graphs of $\langle p_{x} \rangle^{(j)}$ (red solid line), $\langle (p_{x}- \langle p_{x} \rangle^{(j)})^{2} \rangle^{(j)}$ (blue dashed line), $\langle p_{y} \rangle^{(j)}$ (green dotted line) and $\langle (p_{y}- \langle p_{y} \rangle^{(j)})^{2} \rangle^{(j)}$ (purple dashed-dotted line) as functions of $x$. 
(In this paper, any quantity $X^{(j)}$ defined in the region $W_{j}$ for $j=1,2,\cdots,\eta$ is plotted as a function of $x$ so that it takes a constant value $X^{(j)}$ in $ x \in ( (j-1) \Delta L_{x}, j \Delta L_{x})$ for each value of $j$.)
It is shown in this figure that the average $\langle p_{x} \rangle^{(j)}$ of the momentum component $p_{x}$ increases as the position goes away from the boundary with the particle reservoir, while the average $\langle p_{y} \rangle^{(j)}$ of the momentum component $p_{y}$ is zero at any position $x$.

It may be noted that for the ideal gases consisting of point particles realized in the limit of $r\rightarrow +0$ of the hard-disk systems, the average $\overline{p}_{x}$ of the momentum component $p_{x}$ of particles inside the tube is given by $\overline{p}_{x} = \int_{0}^{+\infty} dp_{x}\; p_{x} \sqrt{2\beta/(\pi m)} \exp[-\beta p_{x}^{2}/(2m)]= \sqrt{2 m k_{B}T/\pi}$, which takes the value $\overline{p}_{x} \approx 0.798$ in the case of $m=1$ and $k_{B}T=1$. 
Figure \ref{Fig11PxPyAveFluNoneq} with this fact shows that the average momentum $\langle p_{x} \rangle^{(j)}$ in the $x$-direction is smaller than the value $\overline{p}_{x}$ near the reservoir (for small $x$), and it is larger than $\overline{p}_{x}$ far from the reservoir (for large $x$). 
This feature suggests that averagely speaking, disks are accelerated in the $x$-direction by collisions with other disks injected from the particle reservoir, although the average momentum $\langle p_{x} \rangle^{(j)}$ near the reservoir is smaller than that of the corresponding ideal gas because there would be more number of disks moving toward the boundary $\ca{S}$ by disk-disk collisions in regions closer to $\ca{S}$.

The variances of the momentum components also depend on the $x$-component of position inside the tube. 
We discuss this property by using the quantities $T_{x}^{(j)}$ and $T_{y}^{(j)}$ defined by $T_{x}^{(j)} \equiv \langle (p_{x}- \langle p_{x} \rangle^{(j)})^{2} \rangle^{(j)}/(k_{B}m)$ and $T_{y}^{(j)} \equiv \langle (p_{y}- \langle p_{y} \rangle^{(j)})^{2} \rangle^{(j)}/(k_{B}m)$, for the variances of the momentum components in the $x$- and $y$-directions, respectively. 
Here, the quantity $k_{B} T_{x}^{(j)}/2 = \langle (p_{x}- \langle p_{x} \rangle^{(j)})^{2} \rangle^{(j)}/(2m)$ ($k_{B}T_{y}^{(j)}/2 =  \langle (p_{y}- \langle p_{y} \rangle^{(j)})^{2} \rangle^{(j)}/(2m)$) can be regarded as the average kinetic energy in the $x$-direction (the $y$-direction) in the frame moving with the average velocity of the particle current (with the velocity $(\langle p_{x} \rangle^{(j)}/m,\langle p_{y} \rangle^{(j)}/m)$ in the region $W_{j}$), so we call the quantity $T_{x}^{(j)}$ ($T_{y}^{(j)}$) the 'kinetic temperature' in the $x$-direction (the $y$-direction) in this paper. 
With the kinetic temperatures, the averages $\langle p_{x}^{2}/(2m)\rangle^{(j)}$ and $\langle p_{y}^{2}/(2m)\rangle^{(j)}$ of the kinetic energy $p_{x}^{2}/(2m)$ and $p_{y}^{2}/(2m)$ in the region $W_{j}$ are given by $\langle p_{x}^{2}/(2m)\rangle^{(j)} = k_{B}T_{x}^{(j)}/2 + \langle p_{x}\rangle^{(j)}{}^{2}/(2m)$  and $\langle p_{y}^{2}/(2m)\rangle^{(j)} = k_{B}T_{y}^{(j)}/2 + \langle p_{y}\rangle^{(j)}{}^{2}/(2m)$, respectively. 

In terms of these kinetic temperatures, Fig. \ref{Fig11PxPyAveFluNoneq} shows that the kinetic temperature $T_{y}^{(j)}$ in the $y$-direction decreases as the position goes away from the boundary $\ca{S}$, and the kinetic temperature $T_{x}^{(j)}$ in the $x$-direction is also a decreasing function of $x$ for large values of $x$. 
(It may be noted in Fig. \ref{Fig11PxPyAveFluNoneq} that the kinetic temperature $T_{x}^{(j)}$ is rather an increasing function of $x$ for the regions of $W_{1}$ and $W_{2}$ in which the average $\langle p_{x} \rangle^{(j)}$ of the momentum component $p_{x}$ is smaller than the value of $\overline{p}_{x}$.) 
These kinetic temperatures, especially $T_{x}^{(j)}$, are lower than the temperature $T$ of the particle reservoir, even in the region $W_{1}$ closest to the reservoir. 
This gap between the temperatures $T_{x}^{(j)}$ and $T$ occurs even in the ideal-gas cases. 
Actually, in the ideal gases realized in the limit of $r\rightarrow +0$ of the hard-disk systems, the corresponding kinetic temperature $\ca{T}_{x}$ in the $x$-direction are given by $\ca{T}_{x} =  \int_{0}^{+\infty} dp_{x} (p_{x} - \overline{p}_{x})^{2} \sqrt{2\beta/(\pi m)} \exp[-\beta p_{x}^{2}/(2m)] /(k_{B}m)= [1-(2/\pi)]T$, which is lower than the temperature $T$ of the reservoir. 
In this case, the inequality $\ca{T}_{x} < T$ indicates that the average kinetic energy by particle movements in the $x$-direction consists not only of the energy $k_{B}\ca{T}_{x}/2$ by the kinetic temperature, but also of the energy $\overline{p}_{x}^{2}/(2m)$ by the average particle current.  
It is also important to notice the inequality $k_{B}T_{y}^{(j)}/2 > k_{B}T_{x}^{(j)}/2$ shown in Fig. \ref{Fig11PxPyAveFluNoneq}, suggesting violations of the equipartition law and the local equilibrium hypothesis in the moving frame. 
This feature would imply that values of the particle density are at least not large enough (and there are not enough numbers of particle-particle collisions per unit time) to thermalize particles in local regions. 

\begin{figure}[!t]
\vspfigA
\begin{center}
\includegraphics[width=\widthfig]{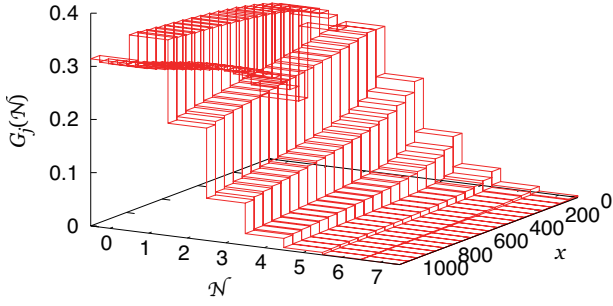}
\vspfigB
\caption{(Color online) 
The probability $G_{j}(\ca{N})$ of the number $\ca{N}$ of particles in the region $W_{j}$ for $j=1,2,\cdots,\eta$ as a function of $\ca{N}$ and $x$, in the model with hard disks in an open tube connected to a particle reservoir. 
}
\label{Fig12ParNumDistr}
\end{center}
\vspfigC
\end{figure}
%
\begin{figure}[!t]
\vspfigA
\begin{center}
\includegraphics[width=\widthfigB]{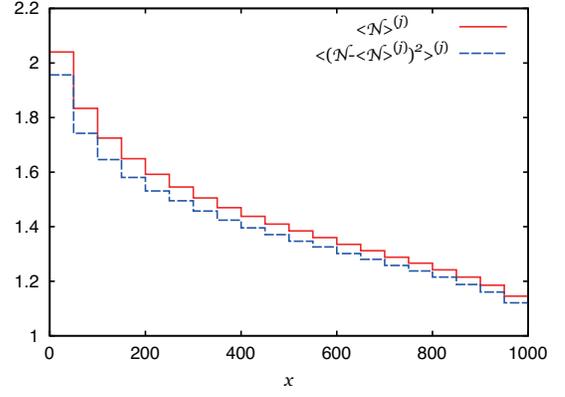}
\vspfigB
\caption{(Color online) 
The average $\langle\ca{N}\rangle^{(j)}$ (red solid line) and the variance $\langle(\ca{N}-\langle\ca{N}\rangle^{(j)})^{2}\rangle^{(j)}$ (blue dashed line) of the number $\ca{N}$ of particles in the region $W_{j}$ for $j=1,2,\cdots,\eta$ as functions of $x$, in the model with hard disks in an open tube connected to a particle reservoir. 
}
\label{Fig13ParNumAveFlu}
\end{center}
\vspfigC
\end{figure}
%
We proceed to discuss local properties related to numbers of disks in our nonequilibrium model for a particle current escaping from a particle reservoir. 
In Fig. \ref{Fig12ParNumDistr} we show the graph of the probability $G_{j}(\ca{N})$ of the number $\ca{N}$ of particles in the region $W_{j}$ for $j=1,2,\cdots,\eta$ as a function of $\ca{N}$ and $x$. 
This figure shows that the probabilities $G_{j}(0)$ and $G_{j}(1)$ are increasing functions of $x$, while $G_{j}(3), G_{j}(4), \cdots, G_{j}(7)$ are decreasing functions of $x$. 
The average $\langle\ca{N}\rangle^{(j)}$ and the variance $\langle(\ca{N}-\langle\ca{N}\rangle^{(j)})^{2}\rangle^{(j)}$ of the number of particles calculated by the probability $G_{j}(\ca{N})$ in the region $W_{j}$ for $j=1,2,\cdots,\eta$ are shown in Fig. \ref{Fig13ParNumAveFlu} as functions of $x$. 
In this figure we see that the average and the variance of the number of particles decrease as the position goes away from the  boundary $\ca{S}$. 
Moreover, the value of $\langle(\ca{N}-\langle\ca{N}\rangle^{(j)}))^{2}\rangle^{(j)}$ is smaller than that of $\langle\ca{N}\rangle^{(j)}$ for any region of $W_{j}$ in this figure, meaning that the probability $G_{j}(\ca{N})$ of the number $\ca{N}$ of particles is not a Poisson distribution (with the average equal to the variance) locally in space.　

\begin{figure}[!t]
\vspfigA
\begin{center}
\includegraphics[width=\widthfig]{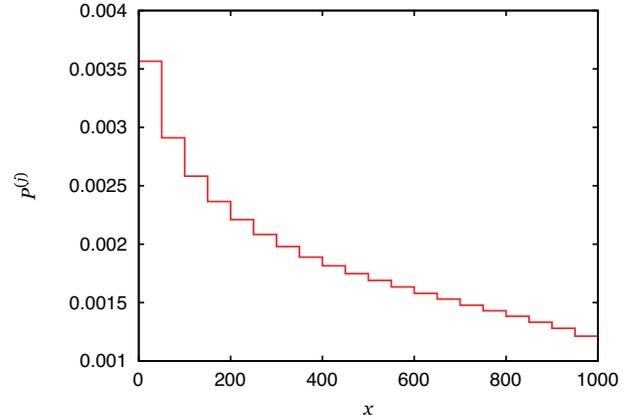}
\vspfigB
\caption{(Color online) 
The pressure $P^{(j)}$ on the wall of the region $W_{j}$ of the tube for $j=1,2,\cdots,\eta$ as a function of $x$, in the model with hard disks in an open tube connected to a particle reservoir.}
\label{Fig14PressNonequil}
\end{center}
\vspfigC
\end{figure}  
%
We calculate the pressure $P^{(j)}$ as the absolute value of the time-average impulse by particle collisions with a wall of the region $W_{j}$ of the tube per unit time and per unit length of the wall for $j=1,2,\cdots,\eta$, and show them in Fig. \ref{Fig14PressNonequil} as a function of $x$. 
It is shown in this figure that the pressure on the wall of the tube decreases as the position goes away from the boundary $\ca{S}$. 
Especially, the pressure $P^{(\eta)}$ near the end $\ca{R}$ of the tube can be less than half as much as the pressure $P^{(1)}$ near the end $\ca{L}$ as the boundary $\ca{S}$, and this decrease ratio of the pressure $P^{(j)}$ from the left end $\ca{L}$ to the right end $\ca{R}$ of the tube is larger than that of the locally average number $\langle\ca{N}\rangle^{(j)}$ of particles shown in Fig. \ref{Fig13ParNumAveFlu}.　

In low-density cases, the average number density of particles moving to the negative $x$-direction is much smaller than that of particles moving to the opposite direction inside the tube because the right end of the tube is open, while particles inside the particle reservoir should move in these both directions with the same probability. 
This implies that the particle number density inside the tube, as well as the pressure on walls of the tube, are much smaller than those inside the particle reservoir, even in a region close to the reservoir. 
For example, in the limit of $r \rightarrow +0$, i.e., in ideal-gas cases, there is no particle moving to the negative $x$-direction inside the tube, so that the particle number density and the pressure by particles inside the tube are half as much as those inside the reservoir. 
Moreover, in such ideal-gas cases, the local momentum probability distributions, the local particle-number density and the local pressure, etc., are independent of $x$. 
In this sense, $x$-dependences of local quantities such as $F_{x}^{(j)}(p_x)$, $F_{y}^{(j)}(p_{y})$, $\langle\ca{N}\rangle^{(j)}$ and $P^{(j)}$ in the cases of $r > 0$, as shown in this subsection \ref{StatisticalMechanicalPropertiesEscapingCurrents}, appear by the particle-particle interactions.

\section{Conclusions and remarks} 
\label{ConclusionsRemarks}

In this paper, we discussed particle-injecting stochastic boundary (PISB) conditions for systems coupled to a particle reservoir with the temperature $T$ and the chemical potential $\mu$. 
In order to describe states of particles injected from a particle reservoir, we imposed the boundary condition that particles injected from the particle reservoir via the boundary $\ca{S}$ have the momentum distribution function (\ref{InjecMomenDistr1}) produced by particles with the Maxwell momentum distribution function (\ref{MaxweDistr1}) with the temperature $T$. 
We also assumed that particle injections occur on the boundary $\ca{S}$ stochastically so that the probability distribution function of time interval between two successive injections of particles via the boundary $\ca{S}$ is given by the exponential function (\ref{TimeInterDistr1}). 
Furthermore, we took into account that the average momentum transfer on the boundary $\ca{S}$ by injected particles or leaving particles via  $\ca{S}$ is given from the pressure in an equilibrium state with the temperature $T$ and the chemical potential $\mu$ by Eqs. (\ref{MomenTrans1}) and (\ref{Press1}), leading to the direct connection (\ref{NuPress1}) between the frequency of the distribution (\ref{TimeInterDistr1}) and the pressure of the particle reservoir. 
Based on these arguments, we constructed the PISB conditions with the external parameters $T$ and $\mu$ in many-particle systems.

In order to check the accuracy of the PISB method, we applied it to two different systems whose equilibrium quantities can be calculated analytically based on grand canonical distributions: two-dimensional ideal gases and low-density hard-disk systems. 
We showed that in these equilibrium systems with the PISB conditions the numerical results for the momentum distribution functions $F_{x}(p_{x})$ and $F_{y}(p_{y})$, the average number $N$ of particles, the pressure $P$ and so on, almost coincide with those calculated analytically from the corresponding grand canonical distributions.  
The well-known equilibrium equations of state, $PV=Nk_{B}T$ for ideal gases and $P(V-2\pi r^{2}N) \approx Nk_{B}T$ for low-density hard-disk systems, were also shown numerically by the PISB method.

The PISB method can be used to investigate equilibrium states described by grand canonical distributions, so it would be meaningful to mention its relations with other methods to produce grand canonical ensembles. 
One of the numerical methods for grand canonical ensembles is the grand canonical Monte Carlo method, in which creations, motions and destructions of particles are imposed inside investigated systems \cite{T99,FS02,AT87}. 
Another example is the grand molecular dynamics method with extended Lagrangian or Hamiltonian dynamics, including a variable governing the dynamics of variation of number of particles \cite{CP91}. 
These methods are proposed to produce grand canonical ensembles in terms of a temperature and a chemical potential, but physical meanings of the dynamics themselves appearing in these methods are unclear. 
In contrast to these methods, the PISB approach uses the Hamiltonian dynamics for particles inside investigated systems, so that it allows to discuss, not only equilibrium states, but also the dynamical processes such as relaxation processes to an equilibrium state or an nonequilibrium steady state.

An advantage of using stochastic boundary conditions for the systems coupled to thermal reservoirs is their applicability to nonequilibrium steady state phenomena.  
To show this feature in the PISB method, in this paper we applied this method to a nonequilibrium model with a particle current escaping from a particle reservoir in an open tube. 
In this model, hard disks can leave from the tube, not only via the boundary $\ca{S}$ located at an end of the tube but also via another end of the tube with an open boundary condition, and a steady particle current escaping from the particle reservoir to the open boundary of the tube is sustained after a long time. 
We discussed statistical-mechanical properties of various quantities in this nonequilibrium model, such as the particle current density as a function of the chemical potential $\mu$, the probability density functions of momentum components of particles, and the particle number density and the pressure as functions of the spatial position in the direction parallel to the tube. 
Many local quantities in this model depend on the spatial position in the tube, differently from equilibrium states, and this model also shows deviations from properties based on the local equilibrium hypothesis.

Particle injections have been widely used to investigate dynamical properties of open systems, known as scattering theoretical approaches. 
In the condensed matter physics, for example, transmission probabilities of injected particles through systems, calculated by the quantum scattering theory, are related to conductivities of the systems \cite{D95,I97,LJ13}. 
In another example, the chaotic systems (e.g. the Lorentz gas model and the multi-baker map model) with the boundary conditions by uniform injecting particle fluxes are considered to discuss nonequilibrium steady states \cite{Gas98}. 
In such open chaotic systems, nonequilibrium invariant measures are introduced, so that some thermodynamical properties, such as Fick's law and the entropy production, can be discussed. 
Generalizations of these arguments by using the PISB conditions including effects of particle-particle interactions and thermodynamic external parameters would be interesting future problems.

Since the PISB method was applied  to ideal gases and hard-disk systems in this paper, now we remark on its applications to other systems such as the systems with soft potentials. 
In contrast to the hard-disk systems in which particle orbits are invariant to a scale change of the total kinetic energy, particle systems with soft potentials do not have such a scale invariance of energy and would show much variety of statistical-mechanical properties.  
Soft potentials also allow us to investigate various types of particles, for example, point particles interacting by the Lennard-Jones potentials \cite{CT80,TC82,MK87,BL91} or molecules \cite{WS12}, for systems coupled to thermal reservoirs. 
For applications of the PISB method to such soft-potential systems, first, we note that the momentum distribution (\ref{InjecMomenDistr1}) of injected particles from a reservoir, as one of the PISB conditions, should be applicable to soft-potential systems, because the equilibrium momentum distribution in Newtonian mechanical systems with Hamiltonians represented as the sum of a kinetic energy and a potential energy [as in Eq. (\ref{HamilNPar1})] is given by the Maxwell momentum distribution, independent of the form of potential energy. 
Actually, the stochastic boundary conditions, in which particles are reflected on a boundary with a coupled heat reservoir by the momentum distribution (\ref{InjecMomenDistr1}) with the temperature $T$ of the heat reservoir, were discussed in many-particle systems with two-body interactions by the Lennard-Jones potentials \cite{CT80,TC82,MK87,BL91}.   
Second, the time-interval distribution (\ref{TimeInterDistr1}) of successive particle injections with the injection frequency (\ref{NuPress1}) would also be applicable to soft-potential systems, as far as successive injections of particles can be regarded to be uncorrelated. 
On the other hand, the PISB approach in soft-potential systems would require to determine stochastically positions of injected particles on the boundary with a coupled reservoir by the equilibrium position distribution based on the soft potential, so that particles are injected from the reservoir more likely at positions with lower potential energies.

It is important to note that the chemical potential $\mu$ in the PISB conditions is introduced via the $\mu$-dependence of the pressure of the particle reservoir, as represented in Eq. (\ref{NuPress1}). 
For ideal gases the analytical form of the pressure as a function of the temperature $T$ and the chemical potential $\mu$ is given by Eq.  (\ref{PressIdeal1}), and such a pressure for hard-disk systems is given by Eq. (\ref{PressDisk1}) in low-density cases.  
However, even in low-density hard-disk cases, there is still a problem to clarify how wide range of values of the chemical potential $\mu$ and the temperature $T$ the form  (\ref{PressIdeal1}) of the pressure is applicable to. 
In the numerical calculations whose results were shown in this paper, we restricted values of the chemical potential $\mu$ to those satisfying the inequality (\ref{RestrChemic1}), but it is only the restriction of $\mu$ in the derivation of Eq. (\ref{PressDisk1}) shown in Appendix \ref{PartiInter}. 
It would also be interesting to investigate statistical-mechanical properties of many-particle systems in high-density cases, with the PISB conditions expressed by their pressures. 
Although pressures of many-particle systems in high-density cases have already been discussed by using canonical distributions \cite{GG84,M08,HM13,S13}, it may be noted that in order to use pressures in the PISB conditions they need to be expressed as functions of $T$ and $\mu$ based on grand canonical distributions.

In the PISB method applied to hard-disk systems, we assumed in this paper that particles are injected randomly via the boundary $\ca{S}$ from a particle reservoir, and there is not a spatial correlation between particles entering to the systems via the boundary $\ca{S}$ and other particles inside the systems, except for the condition in which any injected disk does not overlap with other disks. 
However, spatial correlations of particles by particle-particle interactions would not be negligible in the cases of high-density particles. 
These correlations could occur in many-particle systems, not only with hard-disk interactions, but also with soft-core interactions. 
The formulations of PISB conditions with effects of such spatial correlations in the high-density cases remain as interesting but unsettled problems.

\section*{Acknowledgements} 

One of the authors (T. T.) thanks C. B. McRae and G. P. Morriss for discussing nonequilibrium phenomena with thermal boundary conditions in hard-disk systems.

\appendix
\section{Equilibrium statistical mechanics of many-particle systems based on grand canonical distributions} 

In this appendix, we discuss the analytical expressions of statistical-mechanical quantities of equilibrium systems coupled to a particle reservoir, which are used in the main text of this paper, based on grand canonical distributions. 

\subsection{Equilibrium systems coupled to a particle reservoir} 
\label{AnalyticalEquilibriumProperties}

We consider $d$-dimensional systems described by the Hamiltonian 
\begin{eqnarray}
	H_{\ca{N}}(\bfGamma_{\ca{N}}) = \frac{|\bfp_{\ca{N}}|^{2}}{2m} + U(\bfq_{\ca{N}})
\label{HamilNPar1}
\end{eqnarray}
for $\ca{N}$ identical particles as a function of the phase space vector $\bfGamma_{\ca{N}} = (\bfq_{\ca{N}},\bfp_{\ca{N}})$, where $\bfq_{\ca{N}}$ and $\bfp_{\ca{N}}$ are the position vector and the momentum vector of the $\ca{N}$ number of particles, respectively, $m$ is the mass of each particle, and $U(\bfq_{\ca{N}})$ is the potential energy as a function of $\bfq_{\ca{N}}$. 
The grand partition function of the systems in an equilibrium state with the temperature $T$ and the chemical potential $\mu$ is represented as 
\begin{eqnarray}
   \Xi &=& \sum_{\ca{N}=0}^{+\infty} \Theta_{\ca{N}} e^{\beta \mu \ca{N}} 
\label{GrandParti1}
\end{eqnarray}
with the partition function $\Theta_{\ca{N}}$ of $\ca{N}$ particles, which is given by $\Theta_{0} = 1$ for $\ca{N}=0$ and 
\begin{eqnarray}
   \Theta_{\ca{N}} &=& \int \frac{d\bfGamma_{\ca{N}}}{h^{d\ca{N}}\ca{N}!} 
      e^{-\beta H_{\ca{N}}(\bfsGamma_{\ca{N}})} 
     \nonumber \\
     &=&  \frac{1}{\ca{N}!}\left[\left(\frac{2\pi m}{\beta h^{2}}\right)^{d/2} V_{d}\right]^{\ca{N}} \left(1+\frac{A_{\ca{N}}}{V_{d}^{\ca{N}}}\right)
\label{PartiFunct1}
\end{eqnarray}
for $\ca{N}=1,2,\cdots$, with the volume $V_{d}$ of the spatial area for particles to exist. 
Here, the quantity $A_{\ca{N}}$ is defined by $A_{0} \equiv 0$ for $\ca{N}=0$ and 
\begin{eqnarray}
   A_{\ca{N}} \equiv \int d\bs{q}_{\ca{N}} \; \left[e^{-\beta  U_{\ca{N}}(\bs{q}_{\ca{N}}) } -1 \right]
\label{FunctA1}
\end{eqnarray}
for $\ca{N}=1,2,\cdots$, and includes effects of the potential energy $U_{\ca{N}}(\bs{q}_{\ca{N}})$ in the grand partition function $\Xi$.  
By Eqs. (\ref{HamilNPar1}), (\ref{GrandParti1}) and (\ref{PartiFunct1}) we obtain 
\begin{eqnarray}
   \Xi  &=& \sum\limits_{\ca{N}=0}^{+\infty} \frac{1}{\ca{N}!}\left[\left(\frac{2\pi m}{\beta h^{2}}\right)^{d/2} V_{d} \; e^{\beta\mu}\right]^{\ca{N}} \left(1+\frac{A_{\ca{N}}}{V_{d}^{\ca{N}}}\right) . 
   \nonumber\\
\label{GrandParti2}
\end{eqnarray}
With the grand partition function $\Xi$, the grand potential $J$ is represented as $J = -\beta^{-1}\ln \Xi$, which leads to the entropy $S = -\partial J/\partial T|_{V_{d},\mu}$, the pressure $P = -\partial J/\partial V_{d}|_{T,\mu}$ and the average $N= -\partial J/\partial \mu |_{T,V_{d}} $ of the number $\ca{N}$ of particles, based on the first law of thermodynamics $dJ = -S dT -P dV_{d} -N d\mu$. 

Using the grand partition function $\Xi$, the grand canonical distribution is represented as $\Xi^{-1}\exp\{-\beta [H_{\ca{N}}(\bfGamma_{\ca{N}}) -\mu \ca{N}]\}$. 
For the following arguments, for any quantity $X$ in the systems coupled to a particle reservoir we use the notation $\langle X\rangle$ for the average of the quantity $X$ based on this grand canonical distribution. 
By using the grand partition function $\Xi$ or the grand potential $J = J(T, V_{d},\mu)$ as a function of $T$, $V_{d}$ and $\mu$, the generation function $\tilde{G}(z)$ of the number $\ca{N}$ of particles is represented as   
\begin{widetext}
\begin{eqnarray}
   \tilde{G}(z) &\equiv& \left\langle z^{\ca{N}}\right\rangle = \left\langle e^{\ca{N}\ln z}\right\rangle 
      \nonumber \\
   &=& \Xi^{-1}\sum_{\ca{N}=0}^{+\infty}  \int \frac{d\bfGamma_{\ca{N}}}{h^{d\ca{N}}\ca{N}!}\;   
       \exp\left\{-\beta \left[H_{\ca{N}}(\bfsGamma_{\ca{N}})-\left(\mu + \frac{\ln z}{\beta}\right) \ca{N}\right]\right\}
      \nonumber \\
   &=& \frac{\left.\Xi\right|_{\mu\rightarrow \mu + \beta^{-1}\ln z} }{\Xi}
   = \exp\left\{ -\beta \left[J\!\left(T, V_{d},\mu + \frac{\ln z}{\beta}\right) - J(T, V_{d},\mu)\right]\right\}. 
\label{GenerPartiNumbe1}
\end{eqnarray}
\end{widetext}
We obtain Eq. (\ref{GenerPartiNumbe2}) as Eq. (\ref{GenerPartiNumbe1}) in the case of $d=2$ and $V_{2} = V$. 
By this generation function $\tilde{G}(z)$, the probability distribution $G(\ca{N})$ of the number $\ca{N}$ of particles is given from $G(\ca{N})= (1/\ca{N}!) d^{\ca{N}}\tilde{G}(z)/dz^\ca{N}|_{z=0}$, because of the relation $\tilde{G}(z) = \sum_{\ca{N}=0}^{+\infty} G(\ca{N}) z^{\ca{N}}$. 
It may be noted that the probability distribution $G(\ca{N})$ of the number $\ca{N}$ of particles is also given by
\begin{eqnarray}
   G(\ca{N}) &=& \sum_{K=0}^{+\infty} \int \frac{d\bfGamma_{\ca{N}}}{h^{dK}K!} 
      \;\delta_{K\ca{N}}\; 
      \nonumber \\
   &&\spaEq\spaEq \times
      \Xi^{-1}\exp\left\{-\beta [H_{K}(\bfGamma_{K}) -\mu K]\right\}
      \nonumber \\
   &=& \frac{\Theta_{\ca{N}}}{\Xi}e^{\beta\mu \ca{N}} . 
\label{GenerPartiNumbe4}
\end{eqnarray}
directly from the grand canonical distribution \cite{HM13}.

\subsection{Ideal gases} 
\label{IdelaGasGrand}

We consider the case of ideal gases with no potential energy $U_{\ca{N}}(\bs{q}_{\ca{N}}) = 0$. 
In this case, we obtain $A_{\ca{N}} = 0$ in Eq. (\ref{FunctA1}), leading to the grand partition function (\ref{GrandParti2}) as 
\begin{eqnarray}
   \Xi  = \exp\left[\left(\frac{2\pi m}{\beta h^{2}}\right)^{d/2} V_{d} \; e^{\beta\mu}\right] . 
\label{GrandParti3}
\end{eqnarray}
By Eq. (\ref{GrandParti3}) we obtain the grand potential (\ref{GrandPotenIdeal1}) in the case of $d=2$ and $V_{2}=V$.  
(See Ref. \cite{GN95} for the grand potential of ideal gases in the case of $d=3$.)

Noting that by Eqs. (\ref{PartiFunct1}) and (\ref{GrandParti3}) the partition function $\Theta_{\ca{N}}$ and the grand partition function $\Xi$ satisfy the relations $\Theta_{\ca{N}} \exp(\beta\mu \ca{N}) = \lambda_{d}^{\ca{N}}/\ca{N}!$ and $\Xi = \exp (\lambda_{d})$, respectively, with $\lambda_{d}$ defined by
\begin{eqnarray}
   \lambda_{d} \equiv \left(\frac{2\pi m}{\beta h^{2}}\right)^{d/2}V_{d} \; e^{\beta\mu} 
\label{LambdaDdim1}
\end{eqnarray}
[so $\lambda =\lambda_{2}$ for $\lambda$ given in Eq. (\ref{Lambda2d1})], Eq. (\ref{GenerPartiNumbe4}) leads to Eq. (\ref{PartiNumbeDistrIdeal1}) for the two-dimensional ideal gases.

\subsection{Particle systems with weak interactions} 
\label{PartiInter} 

Now, we consider systems with weak two-body interactions. 
We assume that the potential energy $U_{\ca{N}}(\bs{q}_{\ca{N}})$ is represented as $U_{\ca{N}}(\bs{q}_{\ca{N}}) = 0$ for $\ca{N} = 0,1$ and
\begin{eqnarray}
   U_{\ca{N}}(\bs{q}_{\ca{N}}) = 
       \displaystyle \sumtwo{j=1}{k=1} u\!\left(\left|\bs{q}^{(j)}-\bs{q}^{(k)}\right|\right)  
\label{PotenPartiInter1}
\end{eqnarray}
for $\ca{N} = 2,3,\cdots$ with a function $u(x)$ of a single variable $x$, where $\bs{q}^{(j)}$ is the $d$-dimensional  position vector of the $j$-th particle for $j=1,2,\cdots,\ca{N}$ so $\bs{q}_{\ca{N}} = (\bs{q}^{(1)},\bs{q}^{(2)},\cdots,\bs{q}^{(\ca{N})})$. 
Here, the quantity $u(|\bs{q}^{(j)}-\bs{q}^{(k)}|)$ is the potential energy by the interaction between the $j$-th particle and the $k$-th particle, depending only on the distance $|\bs{q}^{(j)}-\bs{q}^{(k)}|$ between these two particles, for $j \neq k$, $j=1,2,\cdots,\ca{N}$ and $k=1,2,\cdots,\ca{N}$.

We also assume that particle densities of the systems are so low that any simultaneous interaction involving more than three particles is negligible.   
Under this assumption, and by using the potential energy $U_{\ca{N}}(\bs{q}_{\ca{N}})$ defined by $U_{\ca{N}}(\bs{q}_{\ca{N}}) = 0$ for $\ca{N} = 0,1$ and Eq. (\ref{PotenPartiInter1}) for $\ca{N} = 2,3,\cdots$, the quantity $A_{\ca{N}}$ defined by $A_{0} = 0$ for $\ca{N}=0$ and Eq. (\ref{FunctA1}) for $\ca{N} = 1,2,\cdots$ is given by $A_{\ca{N}} = 0$ for $\ca{N} = 0,1$ and is approximated by 
\begin{widetext}	
\begin{eqnarray}
     A_{\ca{N}} &=& 
        \displaystyle \int d\bs{q}_{\ca{N}} \; \left\{\left[\prodtwo{j=1}{k=1} \exp\left[-\beta 
          u\!\left(\left|\bs{q}^{(j)}-\bs{q}^{(k)}\right|\right) \right]\right]-1\right\} 
        \nonumber \\
    &\approx & 
          \displaystyle \frac{\ca{N}(\ca{N}-1)}{2}V_{d}^{\ca{N}-2}\int d\bs{q}^{(j)} \int d\bs{q}^{(k)} \; \left\{\exp\left[-\beta 
          u\!\left(\left|\bs{q}^{(j)}-\bs{q}^{(k)}\right|\right) \right]-1\right\} 
        \nonumber \\
    &\approx & 
       \displaystyle \frac{\ca{N}(\ca{N}-1)}{2}V_{d}^{\ca{N}-1}\int d\bs{x}^{(j,k)} \; \left\{\exp\left[-\beta 
          u\!\left(\left|\bs{x}^{(j,k)}\right|\right) \right]-1\right\} 
        \nonumber \\
    &=& 
        \displaystyle - \ca{N}(\ca{N}-1)V_{d}^{\ca{N}-1} B 
\label{ParamA1}
\end{eqnarray}
\end{widetext}
for $\ca{N} = 2,3,\cdots$, with $B$ defined by
\begin{eqnarray}
   B \equiv \frac{1}{2} \int d\bs{x} \; \left[1-e^{-\beta 
          u (|\bs{x})|)}\right]
\label{InterParamB1}
\end{eqnarray}
as a parameter describing a magnitude of particle-particle interactions. 
Here, the factor $\ca{N}(\ca{N}-1)/2$ is the number of particle pairs chosen from $\ca{N}$ particles, and we introduced $\bs{x}^{(j,k)}$ by $\bs{x}^{(j,k)} \equiv \bs{q}^{(j)}-\bs{q}^{(k)}$, and approximated $\int d\bs{y}$ as $\int d\bs{y} \approx V_{d}$ for $\bs{y} \equiv (\bs{q}^{(j)}+\bs{q}^{(k)})/2$. 
The same approximation as that used to obtain Eq. (\ref{ParamA1}) is also used to describe nonideal gases based on the canonical distribution \cite{R65,LL80}. 
By using the approximation (\ref{ParamA1}) and $\lambda_{d}$ defined by Eq. (\ref{LambdaDdim1}), the grand partition function (\ref{GrandParti2}) is represented as 
\begin{eqnarray}
    \Xi &\approx &\sum_{\ca{N}=0}^{+\infty} \frac{\lambda_{d}^{\ca{N}}}{\ca{N}!}
      -    \frac{B}{V_{d}} \sum\limits_{\ca{N}=2}^{+\infty} \frac{\lambda_{d}^{\ca{N}}}{(\ca{N}-2)!}
      \label{GrandParti4a} 
\end{eqnarray}
approximately. 

We further assume that particle-particle interactions are so weak that for the quantity defined by
\begin{eqnarray}
   \Cd \equiv \frac{B\lambda_{d}}{V_{d}} = B \left(\frac{2\pi m}{\beta h^{2}}\right)^{d/2} \; e^{\beta\mu} ,
\label{WeakInter0}
\end{eqnarray}
the condition 
\begin{eqnarray}
   |\Cd| <\!< 1 
\label{WeakInter1}
\end{eqnarray}
is satisfied. 
Under the condition (\ref{WeakInter1}), the grand partition function (\ref{GrandParti4a}) is given approximately by 
\begin{eqnarray}
    \Xi &\approx &\sum_{\ca{N}=0}^{+\infty} \frac{\lambda_{d}^{\ca{N}}}{\ca{N}!}
      -    \frac{B}{V_{d}} \sum_{\ca{N}=1}^{+\infty} \frac{\lambda_{d}^{\ca{N}+1}}{(\ca{N}-1)!}
      \nonumber \\
   &=& 1+\sum_{\ca{N}=1}^{+\infty} \left(1-\ca{N}\Cd\right)  \frac{\lambda_{d}^{\ca{N}}}{\ca{N}!}
      \nonumber \\
   &\approx& 1+\sum_{\ca{N}=1}^{+\infty} \left(1-\Cd\right)^{\ca{N}}  \frac{\lambda_{d}^{\ca{N}}}{\ca{N}!}
      \nonumber \\
     &=& e^{\lambda_{d}(1- \Cd)} 
\label{GrandParti4}
\end{eqnarray}
where we used the approximation $\left(1-\Cd\right)^{\ca{N}} \approx 1-\ca{N}\Cd$  for any integer $\ca{N}$ under the condition (\ref{WeakInter1}).  
It may be noted that Eq. (\ref{GrandParti4a}) can also be transformed as 
\begin{eqnarray}
   \Xi  &\approx& \sum\limits_{\ca{N}=0}^{+\infty} \frac{\lambda_{d}^{\ca{N}}}{\ca{N}!}
      -    \frac{B}{V_{d}} \sum\limits_{\ca{N}=0}^{+\infty} \frac{\lambda_{d}^{\ca{N}+2}}{\ca{N}!}
      \nonumber \\
    &=&  \left(1- \frac{B\lambda_{d}^{2}}{V_{d}}\right)e^{\lambda_{d}} ,
\label{GrandParti5}
\end{eqnarray}
suggesting the necessary condition 
\begin{eqnarray}
   \frac{B\lambda_{d}^{2}}{V_{d}} &=&  \frac{V_{d}\Cd^{2}}{B} =\lambda_{d}\Cd 
   \nonumber \\
   &=& BV_{d} \left(\frac{2\pi m}{\beta h^{2}}\right)^{d} e^{2\beta\mu} < 1 
\label{RestrChemic1}
\end{eqnarray}
for the derivation of Eq. (\ref{GrandParti5}) because of the positivity $ \Xi > 0$ of the grand partition function $\Xi$.  

By the grand partition function (\ref{GrandParti4}) and Eqs. (\ref{LambdaDdim1}) and (\ref{WeakInter0}), the grand potential $J = -\beta^{-1}\ln \Xi$ is represented by
\begin{eqnarray}
    J &\approx &  -\beta^{-1} \lambda_{d}(1- \Cd)
       \nonumber \\
   &=& - \left(\frac{2\pi m}{ h^{2}}\right)^{d/2}\frac{V_{d} \; e^{\beta\mu}}{\beta^{1+(d/2)}} 
   \left[ 1-  B \left(\frac{2\pi m}{\beta h^{2}}\right)^{d/2} \; e^{\beta\mu}\right]
   \nonumber \\
\label{GrandPoten1}
\end{eqnarray}
approximately.

\subsection{Hard disks with a low density} 
\label{HardDiskGrand}

For two-dimensional systems consisting of hard disks with the radius $r$, the potential energy $u(x)$ for two-disk interactions is expressed as 
\begin{eqnarray}
   u(x) =     \left\{
   \begin{array}{ll}
      +\infty  & \mbox{for $x < 2r$}   \\
       0          & \mbox{for $x > 2r$} .
   \end{array}
   \right. 
\label{PotenParParInt1}
\end{eqnarray}
For this potential with $d=2$, the quantity (\ref{InterParamB1}) is given by  
\begin{eqnarray}
   B  &=& \frac{1}{2} \int_{0}^{+\infty} dR \int_{0}^{2\pi} d\theta \; R \left[1-e^{-\beta u(R)} \right] 
          \nonumber \\
   &=& 2 \pi r^{2} , 
\label{InterParamBDisk1}
\end{eqnarray}
so that the quantity $B$ is twice as large as the area of the disk. 
Inserting Eqs. (\ref{InterParamBDisk1}), $d=2$ and $V_{2}=V$ into Eq. (\ref{GrandPoten1}) we obtain the grand potential (\ref{GrandPotenDisk1}) for the hard-disk systems in low-density cases.


%

\vspace{0.5cm} 
\begin{thebibliography}{99}

\bibitem{R89} H. Risken, 
   \textit{The Fokker-Planck equation: Methods of solution and applications}  
   (Springer-Verlag, Berlin, 1989). 
\bibitem{K92} N. G. van Kampen, 
   \textit{Stochastic processes in physics and chemistry} 
   (Elsevier, Amsterdam, 1992). 
\bibitem{CK12} W. T. Coffey and Y. P. Kalmykov, 
   \textit{The Langevin equation: With applications to stochastic problems 
   in physics, chemistry and electrical engineering} 
   (World Scientific, Singapore, 2012). 
   
   
\bibitem{P10} N. Pottier, 
   \textit{Nonequilibrium statistical physics: linear irreversible processes} 
   (Oxford University Press, Oxford, 2010). 
\bibitem{W08} U. Weiss, 
   \textit{Quantum dissipative systems} 
   (World Scientific, Singapore, 2008). 


\bibitem{CP91} T. \c{C}ag\u{i}n and B. M. Pettitt, 
   Mol. Simul. \textbf{6}, 5 (1991); 
   Mol. Phys. \textbf{72}, 169 (1991). 
\bibitem{T99} J. M. Thijssen, 
   \textit{Computational Physics} 
   (Cambridge University Press, Cambridge, 1999). 
\bibitem{FS02} D. Frenkel and B. Smit, 
   \textit{Understanding molecular simulation: From algorithms to applications} 
   (Academic Press, San Diego, 2002). 
\bibitem{EM08} D. J. Evans and G. Morriss, 
   {\it Statistical mechanics of nonequilibrium liquids} 
   (Cambridge University Press, Cambridge, 2008). 
\bibitem{T20} M. E. Tuckerman, 
   \textit{Statistical mechanics: Theory and molecular simulation} 
   (Oxford University Press, Oxford, 2010). 
\bibitem{LM15} B. Leimkuhler and C. Matthews, 
   \textit{Molecular dynamics: With deterministic and stochastic numerical methods} 
   (Springer International Publishing Switzerland, Cham, 2015). 
\bibitem{A80} H. C. Andersen, J. Chem. Phys. \textbf{72}, 2384 (1980). 
%
\bibitem{PB93} A. Papadopoulou, E. D. Becker, M. Lupkowski, and F. van Swol, J. Chem, Phys. \textbf{98}, 4897 (1993). 
\bibitem{HS94} G. S. Heffelfinger and F. van Swol, J. Chem. Phys. \textbf{100}, 7548 (1994).  


\bibitem{CT80} G. Ciccotti and A. Tenenbaum, J. Stat. Phys. \textbf{23}, 767, (1980). 
\bibitem{TC82} A. Tenenbaum, G. Ciccotti, and R. Gallico, Phys. Rev. A\textbf{25}, 2778 (1982). 
\bibitem{MK84} M. Mareschal and E. Kestemont, Phys. Rev. A\textbf{30}, 1158 (1984). 
\bibitem{MK87} M. Mareschal, E. Kestemont, F. Baras, E. Clementi, and G. Nicolis, Phys. Rev. A\textbf{35}, 3883 (1987). 
\bibitem{BL91} D. K. Bhattacharya and G. C. Lie, Phys. Rev. A\textbf{43}, 761 (1991). 
%
\bibitem{TT98} R. Tehver, F. Toigo, J. Koplik, and J. R. Banavar, Phys. Rev. E\textbf{57}, R17 (1998). 


\bibitem{LS78} J. L. Lebowitz and H. Spohn, J. Stat. Phys. \textbf{19}, 633 (1978). 
\bibitem{BL55} P. G. Bergmann and J. L. Lebowitz, Phys. Rev. \textbf{99}, 578 (1955). 
\bibitem{LF57} J. L. Lebowitz and H. L. Frisch, Phys. Rev. \textbf{107}, 917 (1957). 


\bibitem{PL12} M. Prusty, J. N. Leaw, S. S. Chong, S. A. Cheong, Computer Physics Communications \textbf{183}, 486 (2012). 






\bibitem{D95} S. Datta,  
    \textit{Electronic transport in mesoscopic systems} 
    (Cambridge University Press, Cambridge, 1995). 
\bibitem{I97} Y. Imry, 
   \textit{Introduction to mesoscopic physics} 
   (Oxford University Press, New York, 1997). 
\bibitem{LJ13} M. Lundstrom and C. Jeong, 
   \textit{Near-equilibrium transport: Fundamentals and applications (Lessons from nanoscience: A lecture note series, Vol. 2)} 
   (World Scientific, Singapore, 2013). 

   

\bibitem{CM08} G. Casati, C. Mej\'{i}a-Monasterio, and T. Prosen, Phys. Rev. Lett. \textbf{101}, 016601 (2008). 
\bibitem{HP09} M. Horvat, T. Prosen, and G. Casati, Phys. Rev. E\textbf{80}, 010102R (2009). 


\bibitem{WS12} S. Fritsch, S. Poblete, C. Junghans, G. Ciccotti, L. D. Site, and K. Kremer, 
   Phys. Rev. Lett. \textbf{108}, 170602 (2012); 
   H. Wang, C. Sch\"utte, and L. D. Site, J. Chem. Theory Comput. \textbf{8}, 2878 (2012). 
   
\bibitem{AT87} M. P. Allen and D. J. Tildesley, 
   \textit{Computer simulation of liquids} 
   (Oxford University Press, Oxford, 1987). 
\bibitem{H92} J. M. Haile, 
   \textit{Molecular dynamics simulation: Elementary methods} 
   (Wiley-Interscience, New York, 1992).  


\bibitem{GG84} C. G. Gray and K. E. Gubbins, 
   \textit{Theory of molecular fluids, Volume 1: Fundamentals} 
   (Oxford University Press, Oxford, 1984). 
\bibitem{M08} A. Mulero (Ed.), 
   \textit{Theory and simulation of hard-sphere fluids and related systems} 
   (Springer-Verlag, Berlin, 2008). 
\bibitem{HM13} J.-P. Hansen and I. R. McDonald, 
   \textit{Theory of simple liquids: With applications to soft matter} 
   (Academic Press, Oxford, 2013). 
\bibitem{S13} J. R. Solana, 
   \textit{Perturbation theories for the thermodynamic properties of fluids and solids} 
   (CRC Press, Boca Raton, 2013). 


\bibitem{BB90} W. Bauer and G. F. Bertsch, Phys. Rev. Lett. \textbf{65}, 2213 (1990); 
   O. Legrand and D. Sornette, Phys. Rev. Lett. \textbf{66}, 2172 (1991); 
   W. Bauer and G. F. Bertsch, Phys. Rev. Lett. \textbf{66}, 2173 (1991). 
\bibitem{TS11a} T. Taniguchi and S. Sawada,  
    Phys. Rev. E \textbf{83}, 026208 (2011). 
\bibitem{AP13} E. G. Altmann, J. S. E. Portela, and T. T\'{e}l, 
   Rev. Mod. Phys. \textbf{85}, 869 (2013). 
\bibitem{TS14} T. Taniguchi, H. Murata, and S. Sawada,  
    Phys. Rev. E \textbf{90}, 052923 (2014). 
   
      
\bibitem{K61} R. Kubo, H. Ichimura, T. Usui, N. Hashitsume, 
   \textit{Statistical mechanics: An advanced course with problems and solutions} 
   (Elsevier, Amsterdam, 1965). 


\bibitem{MemoA} 
It may be noted that the position distribution of a hard disk in equilibrium states of many-hard-disk systems is not uniform in general and it has a high probability near hard walls of the systems \cite{ZH96}. 
In this sense, the assumption of a uniform position-distribution of particles injected from the reservoir on the boundary $\ca{S}$ would imply, for example, that the area of the particle reservoir is large enough so that the walls constructing the boundary of the reservoir, which are not perpendicular to the unit vector $\bf{n}$, are far from the boundary $\ca{S}$.   
\bibitem{ZH96} Z. Zheng, G. Hu, and J. Zhang,  Phys. Rev. E\textbf{53}, 3246 (1996). 


\bibitem{Gas98} 
   P. Gaspard, Physica A \textbf{240}, 54 (1997); 
   P. Gaspard, {\it Chaos, scattering and statistical mechanics} (Cambridge University press, Cambridge, 1998). 

\bibitem{GN95} W. Greiner, L. Neise, and H. St\"{o}cker, 
   \textit{Thermodynamics and statistical mechanics} 
   (Springer-Verlag, New York, 1995). 
\bibitem{R65} F. Reif, 
   \textit{Fundamentals of statistical and thermal physics} 
   (McGraw-Hill, Boston, 1965). 
\bibitem{LL80} L. D. Landau and E. M. Lifshitz, 
   \textit{Statistical physics, Part 1: Course of theoretical physics, Volume 5} 
   (Butterworth-Heinemann, Oxford, 1980). 

   
   
\end{thebibliography}
\end{document}